%% file: main.tex
\documentclass[conference]{IEEEtran}

\usepackage{hyperref}
\usepackage{graphicx}
\usepackage{tikz}
\usepackage{sourcecodepro}
\usepackage{comment}
\usepackage{amsthm}
\usepackage[numbers]{natbib}
\usetikzlibrary{shapes, automata, shadows, arrows, positioning}

\tikzstyle{circ} = [draw, rectangle, rounded corners,  text centered, fill= white!20, text height=0.5em, text width=1.5em]
\tikzstyle{circbigger} = [draw, rectangle, rounded corners,  text centered, fill= white!20, text height=0.5em, text width=2.0em]
\tikzstyle{circinvisible} = [draw=white, circle, fill= white!0, text centered, fill= white!20, text height=1.7em, text width=2.5em, inner sep=0pt]
\tikzstyle{line} = [draw, -latex]
\tikzstyle{arrow} = [thick,->,>=stealth]
\tikzstyle{process} = [rectangle, minimum width=3cm, minimum height=1cm, text centered, draw=black, fill=orange!30]
\tikzstyle{nodePrec} = [rectangle, minimum width=3.5cm, minimum height=0.7cm, text centered, draw=black, fill=white!30]

\usepackage{caption}
\def\BibTeX{{\rm B\kern-.05em{\sc i\kern-.025em b}\kern-.08em
    T\kern-.1667em\lower.7ex\hbox{E}\kern-.125emX}}

\input{AGTdefinitions/definitions.sty}
\input{definitions}
\input{defs}
\input{override-defs}
\usepackage{cleveref}

\definecolor{eminence}{RGB}{108,48,130}
\usepackage{stackengine}
\lstdefinelanguage{scala}{
    morekeywords={let,in,if,then,else,fun,def},
    otherkeywords={=>,!,:=,+},
    sensitive=true,
    morecomment=[l]{//},
    morecomment=[n]{/*}{*/},
    morestring=[b]",
    morestring=[b]',
    morestring=[b]""",
    escapeinside={(*}{*)},
    moredelim=**[is][{\btHL}]{`}{`},
  }
\lstdefinelanguage{gsoul}{
  keywords=[1]{def,fn,if,then,else,let,res,try,catch},
  morecomment=[l]{//},
}

\makeatletter
\def\smallunderbrace#1{\mathop{\vtop{\m@th\ialign{##\crcr
   $\hfil\displaystyle{#1}\hfil$\crcr
   \noalign{\kern3\p@\nointerlineskip}%
   \tiny\upbracefill\crcr\noalign{\kern3\p@}}}}\limits}
\makeatother

\IEEEoverridecommandlockouts
\begin{document}

\bstctlcite{IEEEexample:BSTcontrol}

\title{Gradual Sensitivity Typing
\thanks{This work was partially funded by the Millennium Science Initiative Program: code ICN17\_002 and ANID FONDECYT Project 11250054.}}

\author{\IEEEauthorblockN{Dami{\'a}n Arquez}
\IEEEauthorblockA{
PLEIAD Laboratory \\
  Computer Science Department (DCC) \\
  University of Chile \& IMFD\\
  Santiago, Chile \\
  darquez@dcc.uchile.cl
} \and
\IEEEauthorblockN{Mat{\'i}as Toro}
\IEEEauthorblockA{
PLEIAD Laboratory \\
Computer Science Department (DCC) \\
  University of Chile \& IMFD\\
  Santiago, Chile \\
  mtoro@dcc.uchile.cl
} \and
\IEEEauthorblockN{\'Eric Tanter}
\IEEEauthorblockA{
PLEIAD Laboratory \\
  Computer Science Department (DCC) \\
  University of Chile \& IMFD\\
  Santiago, Chile \\
  etanter@dcc.uchile.cl
}
}


\maketitle
\thispagestyle{plain}
\pagestyle{plain}


\begin{abstract}
  Reasoning about the sensitivity of functions with respect to their inputs has interesting applications in various areas, such as differential privacy.
  Sensitivity type systems support the modular checking and enforcement of sensitivity, but can be overly conservative
for certain useful programming patterns. Hence, there is value in bringing the benefits of gradual typing to these disciplines in order to ease their adoption and applicability.
  In this work, we motivate, formalize, and prototype gradual sensitivity typing.
  The language GSoul supports both the unrestricted {\em unknown sensitivity} and bounded imprecision in the form of {\em sensitivity intervals}.
  Gradual sensitivity typing allows programmers to smoothly evolve typed programs without any static sensitivity information towards hardened programs with a mix of static and dynamic sensitivity checking.
  Additionally, gradual sensitivity typing seamlessly enables precise runtime sensitivity checks whenever fully static checking would be overly conservative.
%
  We formalize a core of the language GSoul and prove that it satisfies both the gradual guarantees and sensitivity type soundness, known as metric preservation. We establish that, in general, gradual metric preservation is termination insensitive, and that one can achieve termination-sensitive gradual metric preservation by hardening specifications to bounded imprecision.
Furthermore, we show how the latter can be used to formally reason about the differential privacy guarantees of gradually-typed programs.
%
We implement a prototype of GSoul, which provides an interactive test bed for further exploring the potential of gradual sensitivity typing.
  \end{abstract}

\begin{IEEEkeywords}
gradual typing, differential privacy, function sensitivity
\end{IEEEkeywords}

\everymath{\color{EqBlue}}
\input{sections/introduction}
\input{sections/motivation}

\input{sections/semantics_recipe}

\input{sections/new_gradual_semantics}

\input{sections/metatheory}
\input{sections/dp_reasoning}
\input{sections/related_work}
\input{sections/conclusion}



\bibliographystyle{IEEEtranN}
\bibliography{_Bib/strings,_Bib/pleiad,_Bib/bib,references,_Bib/common}

\end{document}

%% file: definitions.tex
\usepackage{array}
\usepackage{amsmath}
\usepackage{extarrows}
\usepackage{lipsum}
\usepackage{tensor}
\usepackage{upgreek}
\usepackage[colorinlistoftodos]{todonotes}

\PassOptionsToPackage{names,dvipsnames}{xcolor}

\newcommand{\store}{s}

\DeclareDocumentCommand{\reftype}{O{\ttt} O{\Int}}{\{\nu: #2~|~ #1\}}

\renewcommand{\ev}[1][]{{\color{evcolor} \varepsilon_{#1}}}
\newcommand{\evp}[1][]{{\color{evcolor} \varepsilon'_{#1}}}

\NewDocumentCommand{\evi}{m}{{\color{evcolor} \evlangle} #1 {\color{evcolor}\evrangle}}

\NewDocumentCommand{\evbind}{O{1}}{\mathit{bind}_{
  \if 1#1 l \else r \fi
}}

\renewcommand{\nred}{~-->~}

\DeclareDocumentCommand{\enred}{O{\store}}{\overset{#1}{\nred}}
\DeclareDocumentCommand{\ered}{O{\store}}{\overset{#1}{\red}}

\NewDocumentCommand{\ctx}{O{}}{\textcolor{SymPurple}{\mathsf{ctx}}_{#1}}
\newcommand{\sctxKW}{\textcolor{SymPurple}{\mathsf{sctx}}}
\NewDocumentCommand{\sctx}{m m}{\sctxKW{[\SG{#1}, #2]}}

\newcommand{\tenv}[1][]{\Gamma_{#1}}

\newcommand{\emptytenv}{\cdot}
\renewcommand{\emptyenv}{\ensuremath{\raisebox{0.15em}{\scalebox{0.4}{\textbullet}}}\xspace}

\newcommand{\sg}{\SG{s}}
\newcommand{\sd}{\SG{d}}
\newcommand{\eq}{%
\mathrel{
 \!
\settowidth{\@tempdima}{--}%
\resizebox{\@tempdima}{\height}{=}%
\!
}
}

\definecolor{RoyalBlue}{HTML}{0071BC}
\definecolor{RedOrange}{HTML}{F26035}
\definecolor{Emerald}{HTML}{00A99D}
\definecolor{YellowOrange}{RGB}{255, 174, 66}

\definecolor{StoreGreen}{named}{RoyalBlue}
\definecolor{SymPurple}{named}{black}
\definecolor{EqBlue}{named}{black}
\definecolor{evcolor}{named}{RedOrange}

\NewDocumentCommand{\SG}{m}{{\color{StoreGreen} #1}}

\newcommand{\rtype}{\textcolor{SymPurple}{\mathbb{R}}}
\newcommand{\btype}{\textcolor{SymPurple}{\mathbb{B}}}

\newcommand{\emptysenv}{\textcolor{StoreGreen}{\varnothing}}

\newcommand{\sens}[1][]{\textcolor{StoreGreen}{s_{#1}}}

\newcommand{\senv}[1][]{\textcolor{StoreGreen}{\Sigma_{#1}}}
\newcommand{\senvp}[1][]{\textcolor{StoreGreen}{\Sigma'_{#1}}}

\NewDocumentCommand{\sensint}{m}{{\color{StoreGreen} {[#1]}}}
\newcommand{\sensval}[1]{{\color{StoreGreen} #1}}
\newcommand{\unknown}{{\color{StoreGreen} ?}}
\NewDocumentCommand{\venv}{O{}}{\gamma_{#1}}
\NewDocumentCommand{\venvp}{O{}}{\gamma'_{#1}}
\NewDocumentCommand{\venvpp}{O{}}{\gamma''_{#1}}
\NewDocumentCommand{\venvppp}{O{}}{\gamma'''_{#1}}
\NewDocumentCommand{\smap}{O{}}{\phi_{#1}}
\NewDocumentCommand{\smapp}{O{}}{\phi'_{#1}}
\NewDocumentCommand{\smappp}{O{}}{\phi''_{#1}}
\NewDocumentCommand{\smapppp}{O{}}{\phi'''_{#1}}
\NewDocumentCommand{\gsmap}{O{}}{\rho_{#1}}
\NewDocumentCommand{\gsmapp}{O{}}{\rho'_{#1}}
\NewDocumentCommand{\gsmappp}{O{}}{\rho''_{#1}}
\NewDocumentCommand{\gsmapppp}{O{}}{\rho'''_{#1}}
\NewDocumentCommand{\gsvmap}{O{}}{\rho_{#1}}
\NewDocumentCommand{\gsvmapp}{O{}}{\rho'_{#1}}
\NewDocumentCommand{\gsvmappp}{O{}}{\rho''_{#1}}
\NewDocumentCommand{\gsvmapppp}{O{}}{\rho'''_{#1}}

\newcommand{\ascr}[1]{\,\textcolor{SymPurple}{::}\,#1}

\NewDocumentCommand{\sncred}{O{} m m}{#2 \xlongrightarrow{#1}_s #3}
\NewDocumentCommand{\scred}{O{} m m}{#2 \xmapsto{\hspace{0.5em}#1\hspace{0.5em}}_s #3}
\NewDocumentCommand{\screds}{O{} m m O{*}}{#2 \xmapsto{\hspace{0.5em}#1\hspace{0.5em}}^{#4}_s #3}

\NewDocumentCommand{\ncred}{O{} m m}{#2 \xlongrightarrow{#1} #3}
\NewDocumentCommand{\cred}{O{} m m}{#2 \xmapsto{\hspace{0.5em}#1\hspace{0.5em}} #3}
\NewDocumentCommand{\creds}{O{} m m O{*}}{#2 \xmapsto{\hspace{0.5em}#1\hspace{0.5em}}^{#4} #3}

\newcommand{\emptycontext}{\square}

\newcommand{\app}{\,}

\NewDocumentCommand{\lclosure}{O{\venv} m O{\gsmap}}{{\color{SymPurple} \langle} #2, #1 {\color{SymPurple} \rangle}}
\newcommand{\substv}[3]{#3{[#1 / #2]}}

\NewDocumentCommand{\itype}{m m}{{#1}{\cdot}{#2}}

\newcommand{\ttype}[2][]{\mathbb{T}_{\def\temp{#1}\ifx\temp\empty
\empty
\else
#1
\fi
}[#2]}
\newcommand{\cnftype}[1]{\mathbb{C}[#1]}
\NewDocumentCommand{\itconfig}{O{\emptyenv} m}{#1 |- #2}
\NewDocumentCommand{\ittyping}{O{\tenv} O{\svenv} m m}{#1;#2 \vdash #3 : #4}
\NewDocumentCommand{\itsemtyping}{O{\tenv} O{\svenv} m m}{#1;#2 \vDash #3 : #4}
\NewDocumentCommand{\itsemtypingTS}{O{\tenv} O{\svenv} m m}{#1;#2 \vDash_{\mathsf{ts}} #3 : #4}

\NewDocumentCommand{\ictyping}{m O{\venv} m}{
	#1 \in \cnftype{#3}
}

\NewDocumentCommand{\venvwellformed}{m O{\gamma} m}{#3 : #1}

\newcommand{\subty}{<:}

\newcommand{\evjust}{\triangleright}

\newcommand{\evinv}[1]{\mathit{i}\mathsf{#1}}
\newcommand{\icod}{\evinv{cod}}

\newcommand{\iinst}{\evinv{inst}}

\newcommand{\idom}{\evinv{dom}}

\NewDocumentCommand{\sty}{O{}}{\tau_{#1}}
\NewDocumentCommand{\styp}{O{}}{\tau'_{#1}}
\NewDocumentCommand{\stypp}{O{}}{\tau''_{#1}}
\NewDocumentCommand{\styppp}{O{}}{\tau'''_{#1}}
\NewDocumentCommand{\gty}{O{}}{\tau_{#1}}
\NewDocumentCommand{\gtyp}{O{}}{\tau'_{#1}}
\NewDocumentCommand{\gtypp}{O{}}{\tau''_{#1}}
\NewDocumentCommand{\gtyppp}{O{}}{\tau'''_{#1}}
\NewDocumentCommand{\cgty}{O{}}{\mathrm{T}_{#1}}
\NewDocumentCommand{\cgtyp}{O{}}{\mathrm{T}'_{#1}}
\NewDocumentCommand{\cgtypp}{O{}}{\mathrm{T}''_{#1}}
\NewDocumentCommand{\cgtyppp}{O{}}{\mathrm{T}'''_{#1}}
\NewDocumentCommand{\cty}{O{}}{T_{#1}}
\NewDocumentCommand{\ctyp}{O{}}{T'_{#1}}
\NewDocumentCommand{\ctypp}{O{}}{T''_{#1}}
\NewDocumentCommand{\ctyppp}{O{}}{T'''_{#1}}

\newcommand{\gsens}[1][]{\textcolor{StoreGreen}{\mathrm{s}_{#1}}}

\newcommand{\gsenv}[1][]{\textcolor{StoreGreen}{{\Sigma}_{#1}}}
\newcommand{\gsenvp}[1][]{\textcolor{StoreGreen}{{\Sigma}'_{#1}}}
\newcommand{\gsenvpp}[1][]{\textcolor{StoreGreen}{{\Sigma}''_{#1}}}

\newcommand{\transsub}{\trans{}}
\renewcommand{\Isub}{\interior{}}

\newcommand{\denv}[1][]{\textcolor{StoreGreen}{\Delta_{#1}}}

\NewDocumentCommand{\relVV}{m m m O{k}}{\mathcal{V}^{#4}_{#1}\llbracket #2!#3  \rrbracket}
\NewDocumentCommand{\relV}{O{\denv} m m m m O{k}}{(#2 ,#3) \in \glrV[#1]{\itype{#4}{#5}}[#6]}
\NewDocumentCommand{\relVS}{O{\denv} m m m O{k}}{(#2 ,#3) \in \glrV[#1]{#4}[#5]}
\NewDocumentCommand{\relVU}{O{\denv} m m m O{\gsmap}}{(#2 ,#3) \in \mathcal{V}_{#1}\llbracket #4 \rrbracket}

\NewDocumentCommand{\relTT}{O{\denv} m m O{\gsmap}}{\mathcal{T}_{#1}\llbracket #2!#3\rrbracket}
\NewDocumentCommand{\relT}{O{\denv} m m m m O{k}}{(#2 ,#3) \in \mathcal{T}^{#6}_{#1}\llbracket \itype{#4}{#5}  \rrbracket}
\NewDocumentCommand{\relTS}{O{\denv} m m m O{k}}{(#2 ,#3) \in \glrT[#1]{#4}[#5]}
\NewDocumentCommand{\relTI}{O{\denv} m m m m}{\relT[#1]{#2}{#3}{#4}{#5}[I]}

\NewDocumentCommand{\inatom}{m m m m m O{} O{} O{\gsmapp}}{(#1 ,#2) \in \mathsf{Atom}\llbracket#3;#4\rrbracket}

\newcommand{\sv}[1][]{v_{#1}}
\newcommand{\svp}[1][]{v'_{#1}}

\NewDocumentCommand{\relG}{O{\denv} m m m O{k}}{(#2 ,#3) \in \mathcal{S}^{#5}_{#1}\llbracket #4 \rrbracket}

\NewDocumentCommand{\Xisubst}{m O{\r} m}{{[#3/\SG{#2}]}#1}

\NewDocumentCommand{\applyG}{m O{\gsmap}}{#2(#1)}

\newcommand{\e}[1][]{e_{#1}}
\newcommand{\ep}[1][]{\e[#1]'}

\NewDocumentCommand{\rx}{m O{}}{\SG{\mathrm{#1}_{#2}}}

\renewcommand{\r}[1][]{\rx{x}[#1]}

\RenewDocumentCommand{\t}{O{} O{}}{t_{#1}}
\NewDocumentCommand{\tp}{O{} O{}}{\t[#1][#2]'}
\NewDocumentCommand{\tpp}{O{} O{}}{\t[#1][#2]''}
\NewDocumentCommand{\tppp}{O{} O{}}{\t[#1][#2]'''}

\NewDocumentCommand{\mutpe}{O{\alpha} m}{\mu#1.#2}

\newcommand{\listingSG}[1]{\lstinline[mathescape]|$\SG{\text{#1}}$|}

%% file: defs.tex
\usepackage{grid-system}
\usepackage{bbding}
\usepackage{enumitem}

\definecolor{tsmp}{HTML}{870035}

\newenvironment{revb}{\bgroup}{\egroup}

\newcommand{\reviv}[1]{{#1}}

\newcommand{\revvi}[1]{{#1}}
\newenvironment{revvib}{\bgroup}{\egroup}

\newcommand{\revcr}[1]{{#1}}

\newcommand{\ppsal}[1]{{#1}}
\newenvironment{ppsalb}{\bgroup}{\egroup}

\newcommand{\hide}[1]{}
\newcommand{\agthide}[1]{}

\newcommand{\black}{\color{black}}

\newcommand{\tyclsPlur}{types\xspace}

\newcommand{\typeclsPlur}{sensitivity types\xspace}

\newcommand{\LamS}{$\lambda_S$\xspace}

\newcommand{\GSoulCore}{\textsc{GS}\xspace}
\newcommand{\GSoulEvCore}{\textsc{GSCheck}\xspace}

\newcommand{\lambdaSens}{\textsc{Soul}\xspace}

\newcommand{\lang}{\textsc{GSoul}\xspace}

\newcommand{\Jazz}{\textsc{Jazz}\xspace}
\newcommand{\Fuzz}{\textsc{Fuzz}\xspace}
\newcommand{\DFuzz}{\textsc{DFuzz}\xspace}
\newcommand{\muFuzz}{$\color{black}\mu$\textsc{Fuzz}\xspace}
\newcommand{\DDuo}{\textsc{DDuo}\xspace}

\newcommand{\Grefr}[1]{\textcolor{EqBlue}{\textsc{($\cgty$#1)}}}
\newcommand{\IGrefr}[1]{\textcolor{EqBlue}{\textsc{(EV#1)}}}
\newcommand{\ELrefr}[1]{\textcolor{EqBlue}{\textsc{(EL#1)}}}

\NewDocumentCommand{\sttyping}{O{\tenv} O{\svenv} O{} m m}{#1; #2 \vdash_{#3} #4\,:\,#5}

\NewDocumentCommand{\ssttyping}{O{\tenv} O{\svenv} O{s} m m}{#1; #2 \vdash_{#3} #4\,:\,#5}

\NewDocumentCommand{\typewfs}{O{\svenv} m}{#1 \vdash #2}

\newcommand{\rloc}[1][]{l_{#1}}

\definecolor{mutRef}{HTML}{960000}

\newcommand{\rstore}[1][]{\textcolor{mutRef}{\mu_{#1}}}

\newcommand{\Ectx}[1][]{E_{#1}}

\newcommand{\nokont}[1][]{\bot}

\newcommand{\svenv}[1][]{\SG{\Omega_{#1}}}

\newcommand{\polyST}[1][\r]{{\color{StoreGreen}\forall #1}}

\NewDocumentCommand{\polyS}{O{\r}}{{\color{StoreGreen}\Lambda #1}}

\newcommand{\ccleq}{\mathrel{\SG{\lesssim}}}
\newcommand{\nccleq}{\mathrel{\SG{\not\ccleq}}}

\newcommand{\effOp}[1]{\mathbin{\SG{\llparenthesis#1\rrparenthesis}}}

\renewcommand{\ccsub}{\lesssim}

\newcommand{\tjoin}{\curlyvee}
\newcommand{\sjoin}{\mathbin{\color{StoreGreen}\tjoin}}

\newcommand{\appS}[1]{\, \SG{[#1]}}

\newcommand{\tmeet}{\curlywedge}
\newcommand{\smeet}{\mathbin{\color{StoreGreen}\tmeet}}

\newcommand{\domm}{\mathsf{dom}}
\newcommand{\codd}{\mathsf{cod}}

\newcommand{\instt}{\mathsf{inst}}

\NewDocumentCommand{\teff}{m m}{\itype{#1}{\SG{#2}}}

\NewDocumentCommand{\muconf}{O{\rstore} m}{#2} 
\NewDocumentCommand{\rextend}{O{\rloc} m}{\textcolor{mutRef}{[} #1 \mathbin{\textcolor{mutRef}{\mapsto}} #2 \textcolor{mutRef}{]}}

\newcommand{\iteleads}{\leadsto}
\NewDocumentCommand{\itelab}{O{\tenv} O{\svenv} O{} m m m}{\sttyping[#1][#2][#3]{#4 \iteleads #5}{#6}}

  \NewDocumentCommand{\slrV}{O{\denv} m}{\mathcal{V}_{#1} \llbracket #2 \rrbracket}
  \NewDocumentCommand{\slrT}{O{\denv} m}{\mathcal{T}_{#1} \llbracket #2 \rrbracket}
  \NewDocumentCommand{\slrG}{O{\denv} m}{\mathcal{G}_{#1} \llbracket #2 \rrbracket}

  \definecolor{globalVarsColor}{HTML}{fc6603}
  \definecolor{localVarsColor}{HTML}{03d3fc}

  \NewDocumentCommand{\glrV}{O{\denv} m O{k}}{\mathcal{\color{EqBlue}V}_{#1;#3} \llbracket #2 \rrbracket}
  \NewDocumentCommand{\glrT}{O{\denv} m O{k}}{\mathcal{\color{EqBlue}E}_{#1;#3} \llbracket #2 \rrbracket}
  \NewDocumentCommand{\glrG}{O{\denv} m O{k}}{\mathcal{\color{EqBlue}S}_{#1;#3} \llbracket #2 \rrbracket}
  \NewDocumentCommand{\glrM}{O{\denv} O{k}}{\mathcal{\color{EqBlue}M}^{#2}_{#1}}

  \NewDocumentCommand{\glrCoreAtom}{O{\denv} m}{\mathsf{Atom}_{#1} \llbracket #2 \rrbracket}
  \NewDocumentCommand{\glrCoreV}{O{\denv} m}{\mathcal{\color{EqBlue}V}_{#1} \llbracket #2 \rrbracket}
  \NewDocumentCommand{\glrCoreT}{O{\denv} m}{\mathcal{\color{EqBlue}E}_{#1} \llbracket #2 \rrbracket}
  \NewDocumentCommand{\glrCoreG}{O{\denv} m}{\mathcal{\color{EqBlue}S}_{#1} \llbracket #2 \rrbracket}
  \NewDocumentCommand{\glrCoreM}{O{\denv}}{\mathcal{\color{EqBlue}M}_{#1}}

  \NewDocumentCommand{\tsglrCoreAtom}{O{\denv} m}{\mathsf{Atom}_{#1} \llbracket #2 \rrbracket}

  \NewDocumentCommand{\tsglrCoreV}{O{\denv} m}{\mathcal{V}^{\mathsf{ts}}_{#1} \llbracket #2 \rrbracket}
  \NewDocumentCommand{\tsglrCoreT}{O{\denv} m}{\mathcal{E}^{\mathsf{ts}}_{#1} \llbracket #2 \rrbracket}
  \NewDocumentCommand{\tsglrCoreG}{O{\denv} m}{\mathcal{S}^{\mathsf{ts}}_{#1} \llbracket #2 \rrbracket}

  \NewDocumentCommand{\tsglrV}{O{\denv} m O{k}}{\mathcal{V}^{\mathsf{ts}}_{#1;#3} \llbracket #2 \rrbracket}
  \NewDocumentCommand{\tsglrT}{O{\denv} m O{k}}{\mathcal{E}^{\mathsf{ts}}_{#1;#3} \llbracket #2 \rrbracket}
  \NewDocumentCommand{\tsglrG}{O{\denv} m O{k}}{\mathcal{S}^{\mathsf{ts}}_{#1;#3} \llbracket #2 \rrbracket}

  \NewDocumentCommand{\tsglrM}{O{\denv} O{k}}{\mathcal{M}^{#2}_{#1}}

  \NewDocumentCommand{\cxLRV}{O{k} m m m}{#2 \approx_{#1} #3 : #4}
  \NewDocumentCommand{\cxLRTl}{O{k} m m m}{(#2) \leq_{#1} (#3) : #4}
  \NewDocumentCommand{\cxLRTr}{O{k} m m m}{(#2) \geq_{#1} (#3) : #4}
  \NewDocumentCommand{\cxLRT}{O{k} m m m}{(#2) \approx_{#1} (#3) : #4}
  \NewDocumentCommand{\cxLRS}{O{k} m m m}{#2 \approx_{#1} #3 : #4}

\newcommand{\secref}[1]{\S\ref{#1}}
\newcommand{\syntaxLabel}[1]{\textit{\color{black} (#1)}}

\DeclarePairedDelimiter\abs{\lvert}{\rvert}%
\DeclarePairedDelimiter\norm{\lVert}{\rVert}%

\makeatletter
\let\oldabs\abs
\def\abs{\@ifstar{\oldabs}{\oldabs*}}
\let\oldnorm\norm
\def\norm{\@ifstar{\oldnorm}{\oldnorm*}}
\makeatother

\NewDocumentCommand{\reval}{m m}{#1 \Downarrow #2}

\newcommand{\bop}{\mathsf{op}\,}
\newcommand{\bopFuncName}[1][\bop]{\delta_{#1}}

\newcommand{\MP}{\mathsf{MP}}
\newcommand{\EMP}{\mathsf{EMP}}
\newcommand{\GMP}{\mathsf{GMP}}
\newcommand{\GMPts}{\mathsf{GMP}_{\mathsf{ts}}}
\newcommand*{\defeq}{\mathrel{\stackrel{\mathsf{def}}{=}}}

\newcommand{\Zrefr}[1]{\textcolor{EqBlue}{\textsc{($S$#1)}}}
\newcommand{\GZrefr}[1]{\textcolor{EqBlue}{\textsc{(GS#1)}}}

\NewDocumentCommand{\semtypingVenv}{m m}{#1 : #2}
\NewDocumentCommand{\semtypingZ}{O{\tenv} O{} m m}{#1 \vDash_{#2} #3 : #4}
\NewDocumentCommand{\semtypingMP}{O{\tenv} m m}{\semtypingZ[#1][S]{#2}{#3}}
\NewDocumentCommand{\semtypingGMP}{O{\tenv} m m}{\semtypingZ[#1][]{#2}{#3}}
\NewDocumentCommand{\semtypingGMPts}{O{\tenv} m m}{\semtypingZ[#1][\mathsf{ts}]{#2}{#3}}
\NewDocumentCommand{\syntypingZ}{O{\tenv} O{} m m}{\ifx#1\emptytenv \else #1 \fi \vdash_{#2} #3 : #4}
\NewDocumentCommand{\syntypingMP}{O{\tenv} m m}{\syntypingZ[#1][S]{#2}{#3}}
\NewDocumentCommand{\syntypingGMP}{O{\tenv} m m}{\syntypingZ[#1][]{#2}{#3}}
\NewDocumentCommand{\semtyping}{O{\tenv} O{\svenv} m m}{#1; #2 \vDash #3 : #4}

\NewDocumentCommand{\venvClose}{O{\denv} m m O{\tenv}}{d_{#4}(#2, #3) \leq #1}
\NewDocumentCommand{\venvCloseTS}{O{\denv} m m O{\tenv}}{d_{#4}^{\mathsf{ts}}(#2, #3) \leq #1}

\newcommand{\ST}{\ {\text{s.t.}}\ }

\newcommand{\Bty}{B}

\newcommand{\tg}[1][]{t_{#1}}

\newcommand{\tdom}{\mathsf{tdom}}
\newcommand{\tcod}{\mathsf{tcod}}

\newcommand{\ttrans}{\Rightarrow}

\newcommand{\getS}{\mathsf{ms}}
\newcommand{\getMon}{\mathsf{mon}}
\newcommand{\evEq}{\mathsf{rtEq}}
\newcommand{\equiprec}{\mathrel{{\sqsupseteq}{\sqsubseteq}}}
\newcommand{\bounded}{\mathsf{bounded}}

\NewDocumentCommand{\tagged}{m O{\tg}}{#1^{#2}}
\NewDocumentCommand{\checktag}{m O{\tg}}{\mathsf{check}(#1, #2)}

\NewDocumentCommand{\tagelab}{O{\tenv} m m m}{#1 \vdash_{GS} #2 \leadsto #3 : #4}

%% file: override-defs.tex
\RenewDocumentCommand{\itelab}{O{\tenv} m m m}{\syntypingGMP[#1]{#2 \iteleads #3}{#4}}

%% file: sections/introduction.tex
\section{Introduction}

Function sensitivity, also called Lipschitz continuity, is an upper bound of how much the output may change given an input perturbation.
More formally, a function $f$ is $s$-sensitive iff for all $x$ and $y$, $\abs{f(x) - f(y)} \leq s \abs{x - y}$.
Sensitivity plays an important role in different areas such as control theory \cite{zames:ieee1996}, dynamic systems \cite{bournezAl:mfcs2010}, program analysis \cite{chaudhuriAl:fse2011} and differential privacy \cite{dworkRoth:fttcs2014}.
For example, a function $f$ can be made differentially private by adding random noise to its output. The added noise has to be sufficient to guarantee privacy but also tight enough to avoid compromising the utility of the private result.
This is usually achieved by calibrating the amount of noise using the sensitivity of the function.

Many approaches have been proposed to reason about sensitivity,
either statically~\cite{reedAl:icfp2010,
gaboardiAl:popl2013,
zhangAl:icfp2019,
winogradcortAl:icfp2017,
nearAl:oopsla2010,
toroAl:toplas2023,
dantoniAl:fpdcsl2013, abuahAl:arxiv2021a},
or dynamically~\cite{abuahAl2021b}.
Handling sensitivity statically as an effect in the sense of type-and-effects systems~\cite{giffordLucassen:lfp1986} has the advantage of providing strong guarantees statically.
However, type-and-effect systems can make certain useful programming patterns tedious or overly conservative.
Combining the advantages of static and dynamic typechecking is a very active area with a long history, but, to the best of our knowledge, sensitivity has not been explored under this perspective.
One prominent approach for combining static and dynamic typechecking is gradual typing \cite{siekTaha:sfp2006}. Gradual typing supports the smooth transition between dynamic and static checking within the same language by introducing imprecise types and consistent relations. Imprecision is handled optimistically during static typechecking, and is backed by runtime checks in order to detect potential violations of static assumptions.
Hence, the programmer is able to choose at a fine-grained level which portions of the program are dynamically checked and which ones are statically checked.
Gradual typing has been studied in many settings such as subtyping~\cite{garciaAl:popl2016,siekTaha:ecoop2007}, references~\cite{hermanAl:hosc10,siekAl:esop2015}, ownership~\cite{sergeyClarke:esop2012}, information-flow security typing~\cite{disneyFlanagan:stop2011,fennellThiemann:csf2013,toroAl:toplas2018} and refinement types \cite{lehmannTanter:popl2017}, among (many) others.

\hide{In particular, \citet{banadosAl:icfp2014,banadosAl:jfp2016} study a general approach to gradualize type-and-effects in the style of~\citet{giffordLucassen:lfp1986}. \citet{toroTanter:oopsla2015} show how the approach can be made practical in the context of polymorphic effects for the Scala language.
Specifically, the approach of Bañados et al. is based on the generic type-and-effect system of~\citet{marinoMillstein:tldi2009}, which consists of considering effects as (sets of) labels; gradual effect checking is then about handling possible set inclusion between imprecise sets. Unfortunately, while this generic approach to effects encompasses a wide range of effects, sensitivity typing does not fit: sensitivity is not a label, but a {\em quantity}, possibly within a {\em range}, as studied in the type theory literature under different forms~\cite{girardAl:tcs1992,dalLagoGaboardi:lics2011,atkey:lics2018,orchardAl:icfp2019}.
Therefore, we cannot simply define gradual sensitivity types as an instance of the framework developed by Bañados et al.}

The fact that sensitivity is a quantity makes static reasoning particularly challenging, and forces simply-typed approaches to over-approximate sensitivity in problematic ways. For instance, unless one considers some form of dependent typing~\cite{gaboardiAl:popl2013}, the sensitivity of a function that recurses over its argument would be deemed infinite~\cite{reedAl:icfp2010}.
Also, the potential for divergence in a language introduces the possibility of different interpretations of sensitivity type soundness, known as \textit{metric preservation}~\cite{reedAl:icfp2010}.
Metric preservation is a hyperproperty that captures the bound on how much the result of two similar computations may change given an input variation.
In the presence of possible divergence, the situation is similar to that of information flow security, where both termination-sensitive and termination-insensitive notions of noninterference have been studied~\cite{heintzeRiecke:popl1998,zdancewic,goguenMeseguer:ieee1982}.
For metric preservation, a {\em termination-insensitive} interpretation says that, if the function terminates on both inputs, then the output differences is bounded~\cite{abuahAl2021b,azevedoDeAmorimAl:popl2017}. A {\em termination-sensitive} interpretation says that, if the function terminates on the first input, then it also terminates on the second, and the output differences is bounded \cite{reedAl:icfp2010,gaboardiAl:popl2013}.
\reviv{\citet{azevedoDeAmorimAl:popl2017} study a termination-insensitive characterization of metric preservation as well as the necessary conditions to establish when two programs behave the same in terms of termination.}

\begin{revvib}

This work studies the integration of gradual typing with sensitivity typing, with the following contributions:
\begin{itemize}
\item We introduce and motivate \lang, a functional programming language with gradual sensitivity typing (\S\ref{sec:in-action}), featuring a latent type-and-effect discipline with explicit sensitivity polymorphism.
  Notably, \lang not only supports the \emph{unknown} sensitivity---which stands for any sensitivity, possibly infinite---but also a \emph{bounded} form of imprecision in the form of \emph{sensitivity intervals}.
  Bounded imprecision provides programmers with a fine-grained mechanism to balance flexibility and static checking.

  \item After reviewing the key concepts of semantic
 sensitivity typing with a core language and a static syntactic typing discipline
 (\S\ref{sec:sensitivity-typing-background}), we introduce \GSoulCore, a core subset of \lang that captures the key elements of gradual sensitivities (\S\ref{sec:gscore}).

 \item We describe the semantic model of \GSoulCore types, formalizing the novel notion of gradual metric preservation, both termination-insensitive and termination-sensitive (\S\ref{sec:semantics-gscore-ev}). This presentation leaves the exact runtime monitoring mechanism abstract.

\item Following the traditional approach of gradual languages, we describe an internal language \GSoulEvCore that features a type-driven runtime monitoring mechanism based on evidences~\cite{garciaAl:popl2016} (\S\ref{sec:gscore-runtime}). We then describe syntactic typing for \GSoulCore via its elaboration to \GSoulEvCore, and establish its metatheory (\S\ref{sec:gscore-to-gscore-ev}): type safety, gradual guarantees~\cite{siekAl:snapl2015}, and most importantly, sensitivity type soundness. Syntactically well-typed \GSoulCore programs satisfy gradual metric preservation, and furthermore, if their sensitivity type imprecision is {\em bounded}, they satisfy termination-sensitive gradual metric preservation.


  \item We discuss the importance of termination-sensitive gradual metric preservation for differential privacy~\cite{dworkRoth:fttcs2014}, describing two variants of core differentially-private mechanisms and algorithms in \lang (\S\ref{sec:dp_reasoning}).
\end{itemize}

We discuss related work in \S\ref{sec:related} and conclude in \S\ref{sec:conclusion}.
The supplementary material provides proofs of all the formal results presented here. 
\revcr{A prototype implementation of \lang is available at \href{https://github.com/pleiad/gsoul-lang}{https://github.com/pleiad/gsoul-lang}.}

\end{revvib}

%% file: sections/motivation.tex
\section{Gradual Sensitivity Typing in Action}
\label{sec:in-action}

We illustrate the limitations of the conservative nature of static sensitivity typing both for recursive functions, and for a practical scenario based on a standard differential privacy algorithm, Above Threshold~\cite{dworkRoth:fttcs2014}.
We then show how gradual sensitivity typing can address these issues, by supporting the whole spectrum between fully dynamic and fully static sensitivity checking.

\paragraph{Limitations of static sensitivity typing}
We start by exploring how recursive functions are dealt with in a static sensitivity discipline, using
\Fuzz~\cite{reedAl:icfp2010} as a reference.
In this language, $\SG{1}$-sensitive functions are given the special type $\tau_1 \multimap \tau_2$.
Dually, functions of unbounded sensitivity are written as usual: $\tau_1 \to \tau_2$.
For a minimal example, consider the following definition of a recursive \lstinline{scale} function (using concrete \Fuzz syntax), which receives a factor and a value to be scaled:

\begin{small}
\begin{lstlisting}[numbers=none]
scale 0 v = 0
scale n v = v + scale (n-1) v
\end{lstlisting}
\end{small}

The only possible \Fuzz type for this function is $\mathbb{N} \to \mathbb{R} \to \mathbb{R}$, which characterizes it as infinitely sensitive in both arguments (as expressed by the use of arrows $\to$, instead of lollipops $\multimap$).
However, notice that the sensitivity on the scaled value---the second argument---actually depends on the value of the factor argument. Because this value is unknown statically, the type system over-approximates the sensitivity of the function as infinite.
\DFuzz~\cite{gaboardiAl:popl2013} tackles this issue via \emph{lightweight dependent types}: \lstinline|scale| can be given the type $\forall i. \mathbb{N}[i] \to \mathbb{R} \multimap_{\SG{i}} \mathbb{R}$, denoting that the function is $\SG{i}$-sensitive in the second argument.
%
However, such dependent types make the type system more complex for users to harness, and do not always capture more precise sensitivities than \Fuzz (\eg~if at the call site of \lstinline|scale| the first argument value is not known statically, infinity remains the only sound option).



Another problematic restriction of static sensitivity typing manifests for recursive function definitions that capture free variables from their surrounding scope. Consider the \Fuzz typing rule for fixpoints~\cite{reedAl:icfp2010}:

\vspace*{-\baselineskip}

\begin{small}
  \begin{mathpar}
    \inferrule*[]{
      \Gamma, f :_{\SG{\infty}} \tau_1 \multimap \tau_2 \vdash e : \tau_1 \multimap \tau_2
    }{
      \SG{\infty} \Gamma \vdash \mathsf{fix}~f.e : \tau_1 \multimap \tau_2
    }
  \end{mathpar}
\end{small}
This rule reports the fixpoint as infinitely sensitive in captured variables, as observable in the conclusion of the rule where $\Gamma$ is scaled by $\SG{\infty}$.
Indeed, there is no way to know statically how many recursive calls will be performed, and therefore how many times the captured variable will be accessed.
\citet{azevedoDeAmorimAl:popl2017} also identify this issue and design a fixpoint rule that provides a better bound for a particular class of recursive constructs, such as exponentially decaying lists, but their rule does not cover the general case.

Over-approximation in static sensitivity typing can also be problematic in many other practical scenarios.
For instance, consider the Above Threshold (AT) algorithm from the differential privacy literature~\cite{dworkRoth:fttcs2014}. AT receives a list of $\SG{1}$-sensitive queries and a threshold, and approximately returns the index of the first query whose result is above the threshold.
Suppose we have a list of functions $f_1, \dots, f_n$ written in \Fuzz, each representing a query over a database.
Additionally, let us assume that each $f_i$ is \emph{semantically} $\SG{1}$-sensitive.
Since a type system is but a syntactic conservative {\em approximation} of the semantics, we can expect each function $f_i$ to have type $\mathsf{DB} \multimap_{\SG{s_i}} \mathbb{R}$, where $\SG{1 \leq s_i}$.\footnote{
  We leave the $\mathsf{DB}$ type abstract, but the reader can think of it as a set of records, where each record is a tuple of values.
}
Building a list of such functions would necessarily give the list a type that accounts for the {\em worst} sensitivity of the given functions: $(\mathsf{DB} \multimap_{\SG{s_{\mathsf{max}}}} \mathbb{R})\, \mathsf{list}$, where $\SG{s_{\mathsf{max}}} = \max(\SG{s_1}, \SG{\dots}, \SG{s_n})$.
If the sensitivity of a single function in the list is over-approximated, the whole list is impossible to use with AT given that the algorithm expects a list of strictly $\SG{1}$-sensitive queries.


Instead of a list, one could use a tuple of functions of type $\mathsf{DB} \multimap_{\SG{s_1}} \mathbb{R} \times \dots \times \mathsf{DB} \multimap_{\SG{s_n}} \mathbb{R}$, thereby retaining the precise sensitivity information of each function in the tuple. However, if we need to project out the tuple given a statically-unknown index, we are back to the same issue:
the minimal sound return type for $\lambda (i: \mathbb{N}). \lambda (Q: \mathsf{DB} \multimap_{\SG{s'_1}} \mathbb{R} \times \dots \times \mathsf{DB} \multimap_{\SG{s'_n}} \mathbb{R}). Q[i]$ is $\mathsf{DB} \multimap_{\SG{s_{\mathsf{max}}}} \mathbb{R}$,
so any $Q[i]$ is deemed $\SG{s_{\mathsf{max}}}$-sensitive by the type system, despite the fact that, when evaluated it is actually a $\SG{s_i}$-sensitive function. Note that this scenario is outside the reach of \DFuzz as well.

\paragraph{Gradual sensitivities} Gradual sensitivity typing as provided in \lang can elegantly address the limitations of static sensitivity typing, by selectively relying on runtime checking whenever desired or needed. In contrast to runtime sensitivity monitoring~\cite{abuahAl2021b}, which completely foregoes static checking, gradual sensitivity typing does allow static guarantees to be enforced where possible.
To support gradual sensitivity typing, sensitivities in \lang are extended with an unknown sensitivity \listingSG{?}, akin to the unknown type of gradual typing~\cite{siekTaha:sfp2006}.
Intuitively, \listingSG{?} represents \textit{any} sensitivity and allows the programmer to introduce imprecision in the sensitivity information, relying on optimistic static checking, backed by runtime checks.

Gradual sensitivity ordering $\ccleq$ is defined by plausibility: an unknown sensitivity \lstinline{?} is plausibly both smaller and greater than any other sensitivity, \ie~$\unknown \ccleq \sens$ and $\sens \ccleq \unknown$, for any sensitivity $\sens$.
Notice that $\ccleq$ is not transitive: $\SG{2 \ccleq \unknown}$ and $\SG{\unknown \ccleq 1}$, but
$\SG{2 \mathbin{\not\!\ccleq} 1}$.
Plausible ordering induces a notion of \emph{consistent subtyping}: 
given a resource \lstinline|r| in scope, the type \lstinline|Number[r] -> Number[?r]| is a consistent subtype of \lstinline{Number[r] -> Number[r]} because $\unknown \ccleq \SG{1}$.
This notion of consistent subtyping is akin to that studied for
other gradual typing disciplines~\cite{siekTaha:ecoop2007,banadosAl:icfp2014,lehmannTanter:popl2017,garciaAl:popl2016}, but here
focused on the sensitivity information conveyed in types.

The flexibility afforded by consistent subtyping is backed by runtime checks at the boundaries between types of different precision, ensuring at runtime that no static assumptions are silently violated.
For instance, revisiting the AT example in \lang, we can give the list of queries the {\em imprecise} type \lstinline[mathescape]|List<[$\SG{\text{r}}$](DB[$\SG{\text{r}}$] -> Number[$\SG{\text{?r}}$])>|, \ie~a list of functions with unknown sensitivity.\footnote{The syntax \lstinline[mathescape]|[$\SG{\text{r}}$](...)|, used inside the list type parameter, creates a resource-polymorphic type. In practice, these binders could be inferred to make the user-syntax more lightweight.}
As $\unknown$ is plausibly smaller than any other gradual sensitivity, the functions in the list can optimistically be passed to AT, without the need to statically comply with the $\SG{1}$-sensitivity restriction.
If one of the functions in the list happens to violate the $\SG{1}$-sensitivity assumption, AT will simply halt with an error.
In \secref{sec:dp_reasoning}, we define a gradual version of AT that actually avoids erroring---even in the presence of queries that are more than $\SG{1}$-sensitive---and prove its differential privacy guarantees.

Gradual sensitivity typing also offers a simple and effective alternative to handle recursive functions in an exact, albeit dynamically-checked, manner.
For recursive functions that capture external variables, the programmer can use the unknown sensitivity to defer the sensitivity check to runtime, where the number of recursive calls can be observed.
Likewise for functions whose sensitivity with respect to an argument depends on the number of recursive calls: consider the \lstinline|scale| function now defined in \lang:

\begin{small}
\begin{lstlisting}[language=gsoul,numbers=none,mathescape]
def scale(n: Number, res v: Number): Number[$\SG{\text{?v}}$] =
  if (n == 0) then 0 else v + scale(n - 1, v);
\end{lstlisting}
\end{small}

The \lstinline[language=gsoul]|res| modifier in the argument \lstinline|v| indicates that it should be tracked as a resource, and thus the return type of \lstinline|scale| can specify its sensitivity with respect to \lstinline|v|.
Since there is no sound static sensitivity type for \lstinline|scale|, we can use the unknown sensitivity to defer the sensitivity check to runtime, by annotating the return value as \lstinline[mathescape]|Number[$\SG{\text{?v}}$]|.
Then, assuming \lstinline{f} requires an argument that is at most $\SG{10}$-sensitive, we obtain the following behavior:

\begin{small}
\begin{lstlisting}[language=gsoul,numbers=none]
f(scale(10, x))  // typechecks, runs successfully
f(scale(11, x))  // typechecks, fails at runtime
\end{lstlisting}
\end{small}

Therefore, gradual sensitivity typing not only allows programmers to choose between static and dynamic checking as they see fit, but it can also accommodate features that are too conservatively handled by the static discipline.

\paragraph{Bounded imprecision} A key idea from the onset of gradual typing~\cite{siekTaha:sfp2006} is that, while the unknown type $?$ stands for any type whatsoever, one can also use {\em partially-imprecise} type information---for instance using type $\Int ->\,?$ to denote any function type whose domain is $\Int$ but codomain is left open. Partially-imprecise types are key to support the smooth refinement of type information in gradual programs~\cite{siekAl:snapl2015}.

Transposing this idea to the world of gradual sensitivities, we support not only
the unknown sensitivity $\unknown$, which denotes any sensitivity, but also bounded intervals such as $\sensint{1,20}$, which denote any sensitivity {\em within a bounded range}.%
\footnote{Syntactically, we can unify all forms of sensitivities with intervals, considering that
a single sensitivity $\sens$ is shorthand for $\sensint{\sens,\sens}$, and that the unknown sensitivity
$\unknown$ is shorthand for $\sensint{0,\infty}$.}

Sensitivity intervals allow programmers to finely trade flexibility for stronger static guarantees, as well as avoiding runtime checks when possible.
For instance, assume a number variable \lstinline|x| in scope that is $\SG{1}$-sensitive in a resource \listingSG{r}.
Suppose that we build a list \lstinline|l| that contains values obtained by scaling \lstinline|x| by different factors, \eg:
\begin{lstlisting}[language=gsoul,numbers=none,mathescape]
let l : List<Number[$\SG{\text{\_}}$]> =
  List(scale(1, x), scale(2, x), scale(3, x));
\end{lstlisting}
The question arises as to which sensitivity to declare in the type of \lstinline|l| (the \listingSG{\_} above).
To illustrate the implications, consider that we apply one of the following functions, which defer in the sensitivity type of their argument:
\begin{lstlisting}[language=gsoul,numbers=none,,escapechar={|}]
f: Number[|\SG{0r}|] -> Unit    g: Number[|\SG{1r}|] -> Unit
h: Number[|\SG{3r}|] -> Unit
\end{lstlisting}
to some index of \lstinline|l|, say \lstinline|0|, \ie~we compute either \lstinline|f(l[0])|, \lstinline|g(l[0])| or \lstinline|h(l[0])|.
We summarize several cases below depending on which sensitivity is declared for \lstinline|l| and which function we choose to apply to the first element of \lstinline|l|:
\begin{center}
\newcommand{\failsS}{\textcolor{YellowOrange}{type error}\xspace}
\newcommand{\failsD}{\textcolor{RedOrange}{runtime error}\xspace}
\newcommand{\runs}{\textcolor{Emerald}{check ok}\xspace}
\newcommand{\runsN}{\textcolor{gray}{no check}\xspace}
\begin{small}
\begin{tabular}{>{\black}c || >{\black}c | >{\black}c | >{\black}c}
sensitivity of \lstinline|l| (\lstinline|_|) & apply \lstinline|f| & apply \lstinline|g| & apply \lstinline|h| \\
\hline
\lstinline[mathescape]|$\SG{\text{3r}}$| & \failsS & \failsS & \runsN \\
\lstinline[mathescape]|$\SG{\text{?r}}$| & \failsD & \runs & \runs \\
\lstinline[mathescape]|$\SG{\text{0..3r}}$| & \failsD & \runs & \runsN \\
\lstinline[mathescape]|$\SG{\text{1..3r}}$| & \failsS & \runs & \runsN \\
\end{tabular}
\end{small}
\end{center}

In a static language like DFuzz~\cite{gaboardiAl:popl2013} in which \lstinline|scale| can be precisely typed, the most specific and sound sensitivity to declare for \lstinline|l| is the maximum sensitivity of the values put in the list, \ie~\listingSG{3r}. In this case, the applications of \lstinline|f| and \lstinline|g| are rejected by the type system---despite the fact that the underlying value (\lstinline|l[0]|) is really $\SG{1}$-sensitive in \listingSG{r}. The application of \lstinline|h| is well-typed, and requires no runtime checking.

If we declare the sensitivity to be unknown \listingSG{?r}, then all applications are optimistically accepted by the gradual type system, given that it is {\em plausible} for the passed value to meet each function's sensitivity requirement. This static plausibility gives rise to a runtime check, which fails in the case of \lstinline|f|. This is in essence similar to fully dynamic sensitivity checking as in DDuo~\cite{abuahAl2021b}.
In particular, note that a runtime check is necessary including for the application of \lstinline|h|. In contrast, if we refine the upper bound on the sensitivity using \listingSG{0..3r}, then the runtime check for the application of \lstinline|h| is not necessary any longer because the sensitivity of the passed value is {\em definitely} valid.
Likewise, if we choose the even more precise interval \listingSG{1..3r} then applying \lstinline|f| is now statically rejected by the type system, because \lstinline|l[0]| is {\em definitely} at least $\SG{1}$-sensitive in \listingSG{r}. This shows that the gradual type system can exploit partially-imprecise information to deduce definite judgments, which are helpful because they can be (a)~reported to users (\eg~in the IDE) and
(b)~exploited by the implementation to avoid unnecessary runtime checks---an objective known as {\em ``pay-as-you-go''}~\cite{siekTaha:sfp2006}.

While our formal treatment (and \lang prototype) of gradual sensitivity typing does not currently deal with optimizations, we uncover a strong semantic argument in favor of using bounded sensitivity intervals:
they make it possible to provide a stronger, termination-sensitive soundness property for gradual sensitivity typing (\secref{sec:tsmp}), which is crucial for reasoning about differential privacy (\secref{sec:dp_reasoning}).

%% file: sections/semantics_recipe.tex
\section{Background: Typing and Sensitivity}
\label{sec:sensitivity-typing-background}

We first recap essential concepts of typing and semantics using a simply-typed lambda calculus with primitive operations, which lie at the heart of the development of this work. We then complement this background with the standard notion of sensitivity type soundness.

\subsection{Semantics of Types for a Core Calculus}

The syntax of the core calculus is the following:

\vspace*{-\baselineskip}

\begin{small}
\begin{align*}
  \Bty   ::= {}& \rtype \mid \btype & \syntaxLabel{base \tyclsPlur}\\
  \gty    ::= {}&
    \Bty \mid
    \gty \rightarrow \gty
    & \syntaxLabel{\tyclsPlur{}}\\
  e ::= {}&
    c \mid \bop \overline{e} \mid
    x \mid \lambda (x : \gty). e \mid e \app e
    & \syntaxLabel{expressions}
\end{align*}
\end{small}
Types $\gty$ are either base types (real numbers or booleans) or function types.
Expressions $e$ are either constants $c$ \ie~literal values of base types (such as $\mathsf{true}$ or $\mathsf{false}$ for type $\btype$), primitive operations $\bop\!$ applied to a list of expressions $\overline{e}$, variables $x$, lambdas $\lambda (x : \gty). e$, or applications $e \app e$.
We write $\reval{e}{v}$ to denote that expression $e$ reduces to a value $v$;
the call-by-value operational semantics of the calculus are standard~\cite{pierce:tapl}.

Semantically, we say that a closed expression $e$ has type $\gty$, written $\semtypingZ[]{e}{\gty}$, if it evaluates to a value of type $\gty$, \ie~a canonical form of the type $\gty$.
We write $\llbracket \gty \rrbracket$ to denote the set of canonical forms of type $\gty$, \eg~$\llbracket \btype \rrbracket = \set{\mathsf{true}, \mathsf{false}}$.
Formally:\footnote{
  Note that this definition is a {\em termination-sensitive} characterization of semantic typing,
  \ie~in which only terminating expressions are valid inhabitants of a type.
  The termination-insensitive alternative has the form $\reval{e}{v} \implies v \in \llbracket \gty \rrbracket$, making any non-terminating expression a valid inhabitant of any type. We come back to termination issues later in this article.
}

\vspace*{-\baselineskip}

\begin{small}
\begin{align*}
\semtypingZ[]{e}{\gty} \quad \defeq \quad \reval{e}{v} \land v \in \llbracket \gty \rrbracket
\end{align*}
\end{small}
As standard in typed calculi, this notion can be extended to open expressions by introducing type environments $\tenv$, which map variables to types.
Then, we may use substitutions $\venv$---also known as value environments---to close an expression by providing values for its free variables.
We write $\semtypingVenv{\venv}{\tenv}$ to denote that the substitution $\venv$ is well-typed with respect to $\tenv$:

\vspace*{-\baselineskip}

\begin{small}
\begin{align*}
  \semtypingVenv{\venv}{\tenv} \quad \defeq \quad \forall x \in \dom(\tenv). \semtypingZ[]{\venv(x)}{\tenv(x)}
\end{align*}
\end{small}
Using substitutions, we define semantic typing for open expressions: an open expression $e$ has type $\gty$ under a type environment $\tenv$ if, given a well-typed substitution $\venv$, the closed expression $\venv(e)$ is of type $\gty$:

\vspace*{-0.5\baselineskip}

\begin{small}
\begin{align*}
  \semtypingZ[\tenv]{e}{\gty} \quad \defeq \quad \forall \venv \ST \semtypingVenv{\venv}{\tenv}. \semtypingZ[]{\venv(e)}{\gty}
\end{align*}
\end{small}
Notice that this {\em semantic} account of typing is independent of any {\em syntactic} type system---\ie~a decidable typechecking procedure---traditionally noted with $\vdash$, instead of $\vDash$. One can consider the standard typing rules for this core calculus~\cite{pierce:tapl}, and prove the {\em soundness} of the syntactic type system: if an expression is syntactically well-typed, then it is semantically well-typed, \ie~$\vdash e : \gty \implies$ \mbox{$\semtypingZ[]{e}{\gty}$}.

\begin{figure*}
  \begin{small}
  \begin{align*}
    %
    (v_1, v_2) \in \glrCoreV{\itype{\Bty}{\senv}} \quad \defeq \quad {}
    &
    v_i \in \llbracket \Bty \rrbracket \land
    d_{\Bty}(v_1, v_2) \leq \senv \cdot \denv\\
    %
    (v_1, v_2) \in \glrCoreV{\itype{((x : \gty[1]) \rightarrow \cty[2])}{\senv}} \quad \defeq \quad {}
    & \forall \senv[1], \svp[1], \svp[2] \ST (v_1', v_2') \in \glrCoreV{\itype{\gty[1]}{\senv[1]}}.
     ({\sv[1] \app \svp[1]}, {\sv[2] \app \svp[2]}) \in \glrCoreT{\substv{\senv[1]}{x}{\cty[2]} \effOp{+} \senv}\\
    (e_1, e_2) \in \glrCoreT{\cty} \quad \defeq \quad {} &
    \reval{e_i}{v_i} \land (v_1, v_2) \in \glrCoreV{\cty}\\
    (\venv[1], \venv[2]) \in \glrCoreG{\tenv} \quad \defeq \quad {}
    &
    \forall x \in \dom(\Gamma). (\venv[1](x),\venv[2](x)) \in \glrCoreV{\itype{\tenv(x)}{x}}
  \end{align*}
  \end{small}

  \caption{Logical relations for semantic sensitivity typing of \LamS}
  \label{fig:basic-static-logical-relation}
\end{figure*}

\subsection{Semantics of Sensitivity}

Let us explore how to reason about the sensitivity guarantee of expressions in this simple calculus.
We start by revisiting the fundamentals of function sensitivity, and then extend this notion to open expressions in our calculus, where free variables can vary independently.


The sensitivity of a function is a measure of how much the output of the function can vary with respect to variations in the input.
For instance, a function $f : \mathbb{R} \to \mathbb{R}$ is said to be $\sg$-sensitive if and only if for any $x, y \in \mathbb{R}$, $\abs{f(x) - f(y)} \leq \sg \cdot \abs{x - y}$.
This notion can be further generalized to functions between arbitrary metric spaces, \ie~types $T$ equipped with a metric $d_T$:

\begin{definition}[Function Sensitivity]
\label{def:general-function-sensitivity}
A function $f: T_1 \to T_2$ is $\sg$-sensitive iff
for any $\sd \in \mathbb{R}$ and $a_1, a_2 \in T_1$
such that $d_{T_1}(a_1, a_2) \leq \sd$,
we have:

\vspace*{-\baselineskip}

\begin{small}
\begin{align*}
  f(a_1) = v_1 \land f(a_2) = v_2 \land d_{T_2}(v_1, v_2) \leq \sg \cdot \sd
\end{align*}
\end{small}
\end{definition}

Following the approach of~\citet{reedAl:icfp2010}, instead of reasoning about the sensitivity of functions, we can reason about the sensitivity of open expressions.
In this setting, $f$ (from Definition~\ref{def:general-function-sensitivity}) is analogous to an open expression $e$ with (possibly several) free variables.
As such, the sensitivity is captured as a vector that maps each free variable to a sensitivity, denoted $\senv$, and written as $\SG{s_1 x_1} + \dots + \SG{s_n x_n}$.
The input distance, previously captured by $d$, is now captured by a map from free variables to distances, denoted $\denv$, \revcr{also written as $\SG{d_1 x_1} + \dots + \SG{d_n x_n}$}.
The predicted output distance, previously computed as $\sg \cdot \sd$, is now computed as the point-wise product of the sensitivity environment and the distance environment, denoted as $\senv \cdot \denv$, \revcr{and defined as $\SG{s_1 \cdot d_1} + \dots + \SG{s_n \cdot d_n}$}.
Finally, the actual inputs (previously $a_1$ and $a_2$) are now captured by two substitutions $\venv[1]$ and $\venv[2]$ that close the expression, and the actual outputs are captured by the results of applying the substitutions, written $\venv[1](e)$ and $\venv[2](e)$.
Proximity between substitutions is defined as follows:

\vspace*{-0.8\baselineskip}

\begin{small}
\begin{align*}
\venvClose[\denv]{\venv[1]}{\venv[2]}[\tenv]
\ \defeq \
 \forall x \in \dom(\tenv). d_{\tenv(x)}(\venv[1](x), \venv[2](x)) \leq \denv(x)
\end{align*}
\end{small}
With everything in place, we now define metric preservation, the sensitivity property for our lambda calculus:

\begin{definition}[Metric Preservation]
\label{def:static-sensitivity-soundness}
Let $\tenv = x_1 : \Bty_1, \dots, x_n : \Bty_n$.
An open expression $e$ satisfies metric preservation, written $\MP(\tenv, e, \Bty, \senv)$, iff for all $\denv$ and $\venv[1], \venv[2]$, such that $\venvClose[\denv]{\venv[1]}{\venv[2]}[\tenv]$:

\vspace*{-\baselineskip}

\begin{small}
\begin{align*}
\reval{\venv[1](e)}{v_1} \land \reval{\venv[2](e)}{v_2} \land d_{\Bty}(v_1, v_2) \leq \senv \cdot \denv
\end{align*}
\end{small}
\end{definition}

\noindent Observe the direct resemblance between Definitions~\ref{def:general-function-sensitivity} and~\ref{def:static-sensitivity-soundness}.

\subsection{Semantics of Sensitivity Typing}

As the first step for designing a sound sensitivity type system, we enrich the types of the core calculus with sensitivity information.
We call this new calculus \LamS.
We start by indexing the types with a sensitivity environment, written $\itype{\gty}{\senv}$.
The goal is to prove that if an open expression $e$ has type $\itype{\gty}{\senv}$, then it is metric-preserving with respect to $\senv$.
The syntax for our \emph{sensitivity types} is then:\footnote{
  A sensitivity type $\itype{\gty}{\senv}$ may also be seen as a type-and-effect, in the sense of \cite{giffordLucassen:lfp1986}, where the sensitivity environment $\senv$ is the effect part.
}

\vspace*{-\baselineskip}

\begin{small}
\begin{align*}
  \sty    ::= {}&
    \Bty \mid
    (x : \sty) \rightarrow \cty &
  \cty  ::= {}&
    \itype{\sty}{\senv}
\end{align*}
\end{small}
Notice that in an arrow $(x : \gty) \rightarrow \cty$, the codomain $\cty$ may use the variable $x$ in its top-level or any nested sensitivity environment.
For example, a $\SG{2}$-sensitive function type on numbers is written $(x : \rtype) \rightarrow \itype{\rtype}{\SG{2x}}$.

We now turn to the semantic model of sensitivity types.
Like prior work, we rely on binary logical relations~\cite{reedAl:icfp2010,abuahAl2021b,toroAl:toplas2023}; the binary nature comes from the fact that metric preservation characterizes the behavior of two runs of an expression.
The relation (Figure~\ref{fig:basic-static-logical-relation}) is defined using two mutually-defined interpretations: one for values $\glrCoreV{\cty}$, and one for expressions $\glrCoreT{\cty}$.
These interpretations are indexed by a distance environment $\denv$, which models the distance between inputs across two different executions, mirroring the definition of $\MP$.

The interpretation of base types $\itype{\Bty}{\senv}$ is straightforward: the values must be valid canonical forms of the base type $\Bty$ (denoted by $\llbracket \Bty \rrbracket$) and their distance must be bounded by the sensitivity environment $\senv$.
Notice that this definition directly resembles the conclusion part of $\MP$.

Function types $\itype{((x: \gty[1]) \rightarrow \cty[2])}{\senv}$ are interpreted inductively: a pair of functions is related if for any pair of related inputs, the outputs are related.
Remarkably, the outputs must be related at the type resulting of:
\begin{enumerate}[leftmargin=*]
  \item Replacing $x$ in the codomain $\cty[2]$ by the actual sensitivity of the arguments, $\senv[1]$.
    A sensitivity environment substitution, $\substv{\senv[1]}{x}{\senv}$, replaces all occurrences of $x$ in $\senv$ by $\senv[1]$.
    For example, $\substv{{\SG{2y} + \SG{z}}}{\SG{x}}{{(\SG{3x} + \SG{4y})}} = {\SG{3}(\SG{2y} + \SG{z}) + \SG{4y}} = {\SG{10y} + \SG{3z}}$.
    Substitution on types is defined recursively.

  \item Adding the function sensitivity $\senv$, which accounts for the variation between the functions themselves.
    The $\effOp{+}$ meta-operator is defined as $(\teff{\sty}{\senv[1]}) \effOp{+} \senv[2] = \teff{\sty}{\senv[1] {+} \senv[2]}$.
    To illustrate the necessity of this operation, consider the expression $x$ of type $\itype{((y : \rtype) \to \itype{\rtype}{\SG{0x}})}{\SG{1x}}$, closed by $\venv[1] = \set{x \mapsto \lambda y. 0}$ and $\venv[2] = \set{x \mapsto \lambda y. 1}$.
    Note how, if the sensitivity environment $\SG{1x}$ is not considered, the resulting type of applying $\venv[i](x)$ to any pair of arguments will be $\itype{\rtype}{\emptysenv}$, but $d_{\mathbb{R}}(0, 1) = \SG{1} \not\leq \emptysenv \cdot \denv = \SG{0}$.
\end{enumerate}

Two closed expressions are related if, when evaluated, their results are related according to the value interpretation of the same type.
Finally, two substitutions are related if they provide related values for each variable in the type environment.

We define {\em semantic sensitivity typing}, $\semtypingMP[\tenv]{e}{\cty}$, to denote that expression $e$ reduces to close enough---as described by $\cty$---values, when closed by two neighboring substitutions:

\begin{definition}[Semantic Sensitivity Typing]
\label{def:static-semantic-sensitivity-typing}
\begin{small}
\begin{align*}
  \semtypingMP[\tenv]{e}{\cty} \quad \defeq \quad {}
  & \forall \denv, \venv[1], \venv[2] \ST  (\venv[1], \venv[2]) \in \glrCoreG{\tenv}
  . {} \\
  & \quad  ({\venv[1](e)}, {\venv[2](e)}) \in \glrCoreT{\cty}
\end{align*}
\end{small}
\end{definition}

We can then prove that semantic sensitivity typing implies metric preservation:

\begin{lemma}[Soundness of Semantic Sensitivity Typing]
\label{lm:static-semantic-sensitivity-soundness}
Let $\tenv = x_1 : \Bty_1, \dots, x_n : \Bty_n$.
\begin{small}
\begin{align*}
  \semtypingMP[\tenv]{e}{\itype{\Bty}{\senv}} \implies
    \MP(\tenv, e, \Bty, \senv)
\end{align*}
\end{small}
\end{lemma}

\begin{figure}
\begin{small}
\begin{mathpar}
  \inferrule*[lab=\Zrefr{const}]{
    c \in \llbracket \Bty \rrbracket
  }{
    \syntypingMP[\tenv]{c}{\itype{\Bty}{\emptysenv}}
  }

  \inferrule*[lab=\Zrefr{op}]{
    \overline{\syntypingMP[\tenv]{e_i}{\cty[i]}} \\
    \bopFuncName[\bop](\overline{\cty[i]}) = \cty
  }{
    \syntypingMP[\tenv]{
      \bop \overline{e_i}
    }{
      \cty
    }
  }

  \inferrule*[lab=\Zrefr{var}]{
    x : \gty \in \tenv
  }{
    \syntypingMP[\tenv]{x}{\itype{\gty}{x}}
  }

  \inferrule*[lab=\Zrefr{lam}]{
    \syntypingMP[\tenv, x : \gty[1]]{e}{\cty[2]}
  }{
    \syntypingMP[\tenv]{\lambda (x : \gty[1]). e}{\itype{((x : \gty[1]) \rightarrow \cty[2])}{\emptysenv}}
  }

  \inferrule*[lab=\Zrefr{app},width=4cm]{
    \syntypingMP{e_1}{
      \itype{((x : \gty[1]) \rightarrow \cty[2])}{\senv}
    } \\
    \syntypingMP{e_2}{
      \itype{\gty[1]}{\senv[1]}
    }
  }{
    \syntypingMP{
      e_1 \app e_2
    }{
      \substv{\senv[1]}{x}{\cty[2]} \effOp{+} \senv
    }
  }

  \inferrule*[lab=\Zrefr{sub}]{
    \syntypingMP{e}{\cty} \\
    \cty \subty \cty'
  }{
    \syntypingMP{e}{\cty'}
  }
\end{mathpar}
\begin{align*}
  \bopFuncName[+](\itype{\rtype}{\senv[1]}, \itype{\rtype}{\senv[2]}) &= \teff{\rtype}{\senv[1] + \senv[2]}\\
  \bopFuncName[*](\itype{\rtype}{\senv[1]}, \itype{\rtype}{\senv[2]}) &= \teff{\rtype}{\infty (\senv[1] + \senv[2])} \\
  \bopFuncName[\leq](\itype{\rtype}{\senv[1]}, \itype{\rtype}{\senv[2]}) &= \teff{\btype}{\infty (\senv[1] + \senv[2])}
\end{align*}
\end{small}

\caption{Syntactic Sensitivity Typing for \LamS}
\label{fig:lambdaS-syntactic-typing}
\end{figure}

As a final step, Figure~\ref{fig:lambdaS-syntactic-typing} presents a {\em syntactic} sensitivity type system, which is sound with respect to metric preservation.
Constants are typed with an empty sensitivity environment, as they are pure, \ie~they do not depend on any variable.
Primitive n-ary operations $\bop$ are given meaning by the function $\bopFuncName$,  which handles the treatment of sensitivity information.
For a few examples, definitions of $\bopFuncName[]$ are given for addition, multiplication, and comparison.
Multiplication scales the resulting sensitivity by infinite, as the operation is not bounded.
Comparison is similar, but the resulting type is boolean; a small variation in its operands can change the resulting boolean value, hence the infinite sensitivity.
Variables are typed getting their type from the type environment, and their sensitivity is the variable itself, \ie~they are $\SG{1}$-sensitive on themselves.
As standard, the body of a lambda is typed under an extended type environment and, same as constants, the sensitivity of a lambda is pure.
For function applications, we mirror the definition of related functions in the logical relation.
Finally, we admit a subsumption rule that allows an expression to be typed with a supertype of its actual type.
Subtyping is naturally induced by the order of sensitivities, \ie~$(x : \rtype) \to \itype{\rtype}{\SG{2x}} \subty (x : \rtype) \to \itype{\rtype}{\SG{3x}}$ because $\SG{2 \leq 3}$.

We can prove that syntactic typing implies semantic typing:

\begin{lemma}[Syntactic Typing Implies Semantic Typing]
\label{lm:syntactic-implies-semantic-typing}
\begin{small}
\begin{align*}
  \syntypingMP[\tenv]{e}{\cty} \implies \semtypingMP[\tenv]{e}{\cty}
\end{align*}
\end{small}
\end{lemma}

Finally, combining the Lemmas \ref{lm:static-semantic-sensitivity-soundness} and \ref{lm:syntactic-implies-semantic-typing}, we establish that syntactic sensitivity typing implies metric preservation:

\begin{theorem}[Soundness of Syntactic Sensitivity Typing]
\label{thm:static-syntactic-sensitivity-soundness}
Let $\tenv = x_1 : \Bty_1, \dots, x_n : \Bty_n$.
\begin{small}
\begin{align*}
  \syntypingMP[\tenv]{e}{\itype{\Bty}{\senv}} \implies
  \MP(\tenv, e, \Bty, \senv)
\end{align*}
\end{small}
\end{theorem}

\subsection{Metric Preservation and Non-Termination}

So far we have worked under the assumption that expressions are terminating.
While this may be true for the \LamS, it is not the case for more expressive languages; certainly not for gradual languages, such as \lang, where the possibility of errors is first-class.
To conclude this section, we adapt $\MP$ to a weaker, albeit more general, notion of metric preservation that accounts for non-termination, $\EMP$:

\begin{definition}[Exceptional Metric Preservation]
\label{def:exceptional-metric-preservation}
Let ${\tenv = x_1 : \Bty_1, \dots, x_n : \Bty_n}$.
An open expression $e$ satisfies exceptional metric preservation, written $\EMP(\tenv, e, \gty, \senv)$, iff for all $\denv$ and $\venv[1], \venv[2]$, such that $\venvClose[\denv]{\venv[1]}{\venv[2]}$:

\vspace*{-\baselineskip}

\begin{small}
\begin{align*}
  &(\reval{\venv[1](e)}{v_1} \land
    \reval{\venv[2](e)}{v_2}) \implies
  d_{\gty}(v_1, v_2) \leq \senv \cdot \denv
\end{align*}
\end{small}
\end{definition}

Notice that this definition is termination-insensitive, as it does not require that the expressions share the same termination behavior.
Naturally, a termination-sensitive variant could be defined by moving $\reval{\venv[2](e)}{v_2}$ to the conclusion of the implication, \ie $\reval{\venv[1](e)}{v_1} \implies
(\reval{\venv[2](e)}{v_2} \land d_{\gty}(v_1, v_2) \leq \senv \cdot \denv)$.
However, this requires a more subtle analysis of the language semantics.
We come back to this issue in \S\ref{sec:tsmp}.

This concludes the presentation of the blueprint we follow in the remainder of this article to establish the semantics and metatheory of {\em gradual} sensitivity typing.

%% file: sections/new_gradual_semantics.tex

\section{\GSoulCore : A Gradual Sensitivity Language}
\label{sec:gscore}

In this section, we introduce \GSoulCore, the core calculus that serves as the formal foundation of \lang.
Figure~\ref{fig:gscore-syntax} shows the syntax of types and expressions in \GSoulCore, which presents 3 remarkable differences from \LamS:
(1) we extend the syntax of types with a new type constructor $\polyST . \cgty$, (2) we introduce gradual sensitivities $\gsens$ and (3) we extend expressions with ascriptions $e \ascr{\cgty}$.
We discuss each of these in turn.

\begin{figure}
\begin{small}
\begin{align*}
  \gsens &::= \sensint{\sens[1], \sens[2]} \qquad
  \text{\black with:} \quad
  \unknown \triangleq \sensint{0, \infty}, \quad \sens \triangleq \sensint{\sens, \sens}\\
\gsenv &::= \gsens \r + \dots + \gsens \r\\
\gty   &::= \Bty \mid \cgty \to \cgty \mid \polyST . \cgty \\
\cgty &::= \itype{\gty}{\gsenv} \\
e &::= c \mid \bop \overline{e} \mid x \mid \lambda (x : \cgty) . e \mid e \app e \mid \polyS . e \mid e \appS{\gsenv} \mid e \ascr{\cgty}
\end{align*}
\end{small}

\caption{Syntax of \GSoulCore}
\label{fig:gscore-syntax}
\end{figure}

\subsection{Distance Variables for Metric Preservation}
\label{sec:distance-variables}

Across \S\ref{sec:sensitivity-typing-background}, we explored the semantics of sensitivity modeling variations through free variables in expressions.
While this is convenient for reasoning about semantics, it may not be practical for a programming language, as it treats all variables as resources worth tracking, which consequently adds a significant amount of information to types.
\lang instead allows the programmer to specify which variables should be tracked as resources, \eg~consider a simple function:

\begin{small}
\begin{lstlisting}[language=gsoul,numbers=none,mathescape]
def foo(a: Number, res b: Number): Number[2b] =
  a + b + b;
\end{lstlisting}
\end{small}

The \lstinline[language=gsoul]{res} modifier in the function declaration specifies that the variable \lstinline[language=gsoul]{b} should be tracked as a resource.
Consequently, the return type can be annotated with the sensitivity of the body with respect to \lstinline[language=gsoul]{b}.
Observe that the first argument \lstinline[language=gsoul]{a} is not tracked for sensitivity, because it is not marked as a resource.

In order to model this specification, let us revisit the syntax of function types for \LamS, $(x : \sty[1]) \to \cty[2]$.
Here, the sensitivity information conveyed in the codomain of the function, $\cty[2]$, is dependent on the name of the argument, $x$, hence the need of specifying the name of the variable in the domain---as in $(x : \sty[1])$.
For \GSoulCore, instead of having all functions treat their argument as resources, we introduce a more general construct that decouples the role of $(x : \_)$ from function types:
$\polyST[\r] . \cty$, which abstracts a sensitivity type $\cty$ over a {\em distance variable} $\r$.
Distance variables $\r$ (written in blue roman typestyle) differ from lambda-bound variables $x$ in that they live at the type level, and are used solely to describe the sensitivity of computations.
Then, the type $(x : \sty[1]) \to \cty[2]$ can be now written as $\polyST[\rx{x}] . \itype{\sty[1]}{\rx{x}} \to \cty[2]$, noting that the syntax of function types is now simply $\cty[1] \to \cty[2]$.
Accordingly, we introduce the distance abstraction term  $\polyS[\r] . e$.
A distance abstraction is eliminated by instantiating the distance variable with a concrete sensitivity environment, using the distance application form $e \appS{\gsenv}$. This is akin to explicit parametric polymorphism~\cite{reynolds:83}.

Note that the \lstinline[language=gsoul]{res} modifier can be implemented simply as syntactic sugar for a distance abstraction.
For instance, the function $\polyS[\r]. \lambda (n : \itype{\mathbb{R}}{\r}). n + n$ is the same as the following function---using concrete syntax and the \lstinline[language=gsoul]{res} modifier:
\begin{lstlisting}[language=gsoul,numbers=none,mathescape]
def double(res n: Number): Number[$\SG{\text{2n}}$] = n + n;
\end{lstlisting}

In fact, the \lstinline[language=gsoul]{res} modifier is implemented as a desugaring step in the parser, which transforms the function into a distance abstraction as follows:
\begin{lstlisting}[language=gsoul,numbers=none,mathescape]
def double[$\SG{\text{x}}$](n: Number[$\SG{\text{1x}}$]): Number[$\SG{\text{2x}}$] = n + n;
\end{lstlisting}

Here, the distance variable \listingSG{x} is introduced and used to parameterize the distance of the input variable \lstinline[language=gsoul]{n}.
Furthermore, when applying a function that was declared using the \lstinline[language=gsoul]{res} modifier, \lang does not require to explicitly eliminate the distance variable using a concrete sensitivity environment, as it is done automatically by using the inferred sensitivity environment from the actual argument (similarly to how a practical polymorphic language like Haskell dispenses the programmer from explicitly providing type applications).

\subsection{Imprecise Sensitivity Information}
\label{sec:gradual-sensitivities}

{\em Gradual sensitivities} represent {\em imprecise} sensitivity information that is optimistically handled by the type system.
A gradual sensitivity $\gsens$ is defined as a valid interval of two static sensitivities, capturing the plausibility of a sensitivity being any number within the range (Figure~\ref{fig:gscore-syntax}).\footnote{
  By extension, types, expressions and type environments may be gradual, though
  we retain the same meta-variables from the syntax presented earlier.
}
Consequently, the unknown sensitivity $\unknown$ is simply sugar for the interval $\sensint{0, \infty}$, and a \textit{fully precise} sensitivity $\sens$ is sugar for the interval $\sensint{\sens,\sens}$.
Precision in the context of gradual sensitivities corresponds to interval inclusion.
Formally, $\sensint{\sens[1], \sens[2]}$ is more precise than $\sensint{\sens[3], \sens[4]}$, written $\sensint{\sens[1], \sens[2]} \gprec \sensint{\sens[3], \sens[4]}$, if and only if $\sens[1] \geq \sens[3]$ and $\sens[2] \leq \sens[4]$.
Then, as one would expect, the unknown sensitivity is the least precise sensitivity and, conversely, a sensitivity $\sensint{\sens, \sens}$ is said to be fully precise, as no other sensitivity is more precise than itself but the same one.

Sensitivity precision naturally induces a notion of precision on types, \eg~$\itype{\rtype}{\sensint{5, 10}\r} \to \itype{\rtype}{\sensint{10, 10}\r} \gprec \itype{\rtype}{\sensint{0, 10}\r} \to \itype{\rtype}{\sensint{5, 15}\r}$, because $\sensint{5, 10} \gprec \sensint{0, 10}$ and $\sensint{10, 10} \gprec \sensint{5, 15}$.
\revcr{Type precision captures the idea of a type containing more precise sensitivity information than another type, hence it corresponds to checking the point-wise precision of the sensitivities in the involved types.
Consequently, type precision in gradual typing is covariant in the domain of arrows~\cite{garciaAl:popl2016,siekAl:snapl2015}.}

Plausibility for gradual sensitivities arises from relaxing the order of sensitivities, which yields an optimistic relation for gradual sensitivities, denoted by $\ccleq$.
Intuitively, a gradual sensitivity $\gsens[1]$ is less than or equal to another, $\gsens[2]$, if there \emph{exists} a sensitivity within $\gsens[1]$ that is less or equal to another one within $\gsens[2]$.
This is easily checkable by comparing the lower bound of $\gsens[1]$ with the upper bound of $\gsens[2]$, which corresponds to the best-case scenario.
Formally, we define:

\begin{definition}[Consistent Sensitivity Ordering]
  \label{def:consistent-sens-subtyping}
  $\sensint{\sens[1], \sens[2]} \ccleq \sensint{\sens[3], \sens[4]}$ if and only if $\sens[1] \leq \sens[4]$.
\end{definition}

To illustrate the plausible nature of this relation, note that $\SG{5} \ccleq \sensint{0, 10}$ holds, even though there are sensitivities within $\sensint{0, 10}$ that are not less than or equal to $\SG{5}$.
Conversely, we still have that $\SG{10} \nccleq \sensint{0,5}$, because there is no sensitivity within $\sensint{0,5}$ that is greater than or equal to $\SG{10}$.
Lastly, notice that this definition satisfies the behavior described in \S\ref{sec:in-action} for the unknown sensitivity: $\unknown \ccleq \gsens$ and $\gsens \ccleq \unknown$ for any $\gsens$.

One crucial aspect of consistent sensitivity order is that, in general, the transitivity of two consistent judgments is not guaranteed to hold.\footnote{
  \revcr{
  Note that the consistent lifting of a transitive relation is not generally transitive~\cite{garciaAl:popl2016}. Therefore, the consistent lifting of a (pre)order is not itself a (pre)order.
  Yet the convention for consistent liftings is to preserve the name of the original lifted relation, qualifying it with ``consistent'' (such as {\em consistent subtyping}~\cite{garciaAl:popl2016,siekTaha:ecoop2007}). So, {\em consistent sensitivity ordering} is not a preorder, just like {\em consistent label ordering} in gradual security typing~\cite{toroAl:toplas2018} is not.}
}
For instance, although $\SG{10} \ccleq \unknown$ and $\unknown \ccleq \SG{5}$, the transitive judgement of both, $\SG{10} \ccleq \SG{5}$, does not hold.
Nevertheless, in some cases the transitivity of two consistent judgments is indeed satisfied, \eg~with $\SG{5} \ccleq \unknown$ and $\unknown \ccleq \SG{10}$, we indeed have that $\SG{5} \ccleq \SG{10}$.
Just like with static sensitivities, the order of gradual sensitivities induces an order on sensitivity environments and types.
We use $\ccsub$ to denote {\em consistent subtyping}, which is the gradual counterpart of subtyping for gradual sensitivities.
For example, $\itype{\rtype}{\unknown\r[1]} \to \itype{\rtype}{\SG{10}\r[2]} \ccsub \itype{\rtype}{\SG{5}\r[1]} \to \itype{\rtype}{\unknown\r[2]}$, because $\SG{5} \ccleq \unknown$ and $\SG{10} \ccleq \unknown$.
 Consistent subtyping is defined in Figure~\ref{fig:consistent-subtyping}.

\begin{figure}
\begin{small}
\begin{mathpar}
  \inferrule*[lab=\Grefr{sub-b}]{ }{
    \Bty \ccsub \Bty
  }

  \inferrule*[lab=\Grefr{sub-arrow}]{
    \cgty[21] \ccsub \cgty[11] \\
    \cgty[12] \ccsub \cgty[22]
  }{
    \cgty[11] \rightarrow \cgty[12] \ccsub
    \cgty[21] \rightarrow \cgty[22]
  }

  \inferrule*[lab=\Grefr{sub-forall}]{
    \cgty[1] \ccsub \cgty[2]
  }{
    \polyST . \cgty[1] \ccsub
    \polyST . \cgty[2]
  }
\end{mathpar}
\end{small}
\caption{Consistent subtyping}
\label{fig:consistent-subtyping}
\end{figure}

\subsection{Ascriptions}
\label{sec:ascriptions}

Lastly, we extend expressions with type ascriptions $e \ascr{\cgty}$, which represent a syntax-directed subsumption assertion.
Well-typedness (either semantic or syntactic) of an ascription, in a gradual setting, is given by the consistent subtyping relation, \eg~if $e$ has type $\cgty[1]$ and $\cgty[1] \ccsub \cgty[2]$, then $e \ascr{\cgty[2]}$ should be well-typed.

Given that consistent subtyping is induced by the consistent sensitivity order, transitivity of two (consistent) subtyping judgments is also not guaranteed, \ie~if $\cgty[1] \ccsub \cgty[2]$ and $\cgty[2] \ccsub \cgty[3]$, then $\cgty[1] \ccsub \cgty[3]$ may not hold.
This behavior is key to allow for plausibility in a gradual language.
Conversely, this lack of general transitivity for consistent subtyping judgments, implies that ascription must be dealt with carefully in the reduction semantics.
Whereas in a static typing setting, type ascriptions only serve the purpose of weakening the type of an expression---thus being a no-op in the reduction semantics---, in a gradual setting, ascriptions may be used to optimistically lower the sensitivity of an expression, which if necessary, in order to avoid violating the sensitivity guarantees of the language, should produce an error during runtime.
To illustrate, suppose $e$ is semantically 2-sensitive on some $\r$ and, consequently, has type $\itype{\rtype}{\SG{2}\r}$.
Then, $e' = e \ascr{\itype{\rtype}{\unknown\r}} \ascr{\itype{\rtype}{\SG{1}\r}}$ should be well-typed, since $\SG{2} \ccleq \unknown$ and $\unknown \ccleq \SG{1}$.
Nonetheless, $e'$ is clearly not 1-sensitive, and thus should produce an error during runtime.
Otherwise, two runs of $e'$ would eventually produce results further than what is described by the sensitivity type.
We present a mechanism to perform such checks in \S\ref{sec:gscore-runtime}.

\section{Semantics of Gradual Sensitivity Typing}
\label{sec:semantics-gscore-ev}

\begin{figure*}
  \begin{small}
  \begin{align*}
    %
    (v_1, v_2) \in \glrCoreV{\itype{\Bty}{\gsenv}} \quad \defeq \quad {}
    &
    v_i \in \llbracket \Bty \rrbracket \land
    d_{\Bty}(v_1, v_2) \leq (\senv[1] \sjoin \senv[2]) \cdot \denv
    \quad \text{\textcolor{black}{where}}~ \senv[i] = \getS(\getMon(v_i))\\
    %
    (v_1, v_2) \in \glrCoreV{\itype{(\cgty[1] \rightarrow \cgty[2])}{\gsenv}} \quad \defeq \quad {}
    & \forall \svp[1], \svp[2], (v_1', v_2') \in \glrCoreV{\cgty[1]}.
     ({\sv[1] \app \svp[1]}, {\sv[2] \app \svp[2]}) \in \glrCoreT{\cgty[2] \effOp{+} \gsenv}\\
    (v_1, v_2) \in \glrCoreV{\itype{(\polyST.\cgty)}{\gsenv}} \quad \defeq \quad {}
    & \forall \gsenvpp. ({\sv[1] \appS{\gsenvpp}}, {\sv[2] \appS{\gsenvpp}}) \in \glrCoreT{\Xisubst{\cgty}[\r]{\gsenvpp} \effOp{+} \gsenv}\\
    (e_1, e_2) \in \glrCoreT{\cgty} \quad \defeq \quad {} &
    (\reval{e_1}{\sv[1]} \land
    \reval{e_2}{\sv[2]}) \implies
    (v_1, v_2) \in \glrCoreV{\cgty}\\
    (\venv[1], \venv[2]) \in \glrCoreG{\tenv} \quad \defeq \quad {}
    &
    \forall x \in \dom(\Gamma). (\venv[1](x),\venv[2](x)) \in \glrCoreV{\Gamma(x)}
  \end{align*}
  \end{small}
  \caption{Logical relations for semantic gradual sensitivity typing}
  \label{fig:gradual-logical-relations}
\end{figure*}

We now define the semantic model for \GSoulCore types for which, just like in \S\ref{sec:sensitivity-typing-background}, we use a binary logical relation.
Before we delve into the details, we write $\reval{e}{v}$ and $\reval{e}{\error}$ to denote that $e$ reduces to the value $v$ or an error, respectively.

\subsection{Distance Reasoning for Gradual Sensitivities}

Let us start by defining $\glrCoreV{\itype{\Bty}{\gsenv}}$.
Whereas in a static setting, the output distance is directly computable (from $\denv$ and $\gsenv$), in a gradual setting, imprecision makes it less obvious how to compare two values.
Moreover, given the optimism introduced by gradual sensitivities, the actual sensitivity of an expression can only be observed after performing the operations that lead to a value.
For instance, the constant $2$ obtained from $x + 1$ is different to that obtained from $x + x$, where $x = 1$; the former is $\SG{1}$-sensitive on $x$, while the latter is $\SG{2}$-sensitive.
Nevertheless, both expressions could be ascribed to $\itype{\rtype}{\unknown\r}$, hiding the actual sensitivity of the expressions in the type.
Strictly speaking, the information conveyed in $\gsenv$ does not provide a precise enough sensitivity to compare two values; it has not been refined by the actual operations performed.
The key insight for defining a logical relation for gradual sensitivity typing, is the necessity of monitoring the sensitivity of expressions during evaluation, and using this information to compare values.

\paragraph*{A monitor for gradual sensitivities}
Based on the operations performed to reach a value $v$ from an initial expression $e$, $v$ should contain the {\em monitored} sensitivity information of $e$.
We do not prescribe here a specific representation of monitors, but instead only specify its interface.
We provide concrete definitions in the next section.
First of all, it should be possible to obtain the monitor of a value, with the metafunction $\getMon$.
\revcr{This monitor should contain (1) any necessary information to perform runtime sensitivity checks, and (2) the actual monitored sensitivity of the computation that produced the value.}
\revcr{Because of the second point,} another metafunction $\getS$ can be used to extract a fully-precise sensitivity environment from a monitor, which should correspond to the least conservative approximation of the actual sensitivity of the initial expression, based on the operations performed to reach such value.
Moreover, we require monitors to be adequate with respect to the initial sensitivity typing information of expressions, in the sense that if $e$ has type $\itype{\gty}{\gsenv}$ and $\reval{e}{v}$, then $\getS(\getMon(v)) \gprec \gsenv$.

\subsection{Logical relations for Gradual Sensitivity Typing}

Figure~\ref{fig:gradual-logical-relations} shows the definition of the logical relation for gradual sensitivity typing.
As described before, for base types, we obtain the monitored sensitivity environments from the values, computing $\senv[i] = \getS(\getMon(v_i))$.
Since the two values may report different sensitivities, we need to join these sensitivities to compute a conservative distance between the values.
Hence, the condition over the distance of values is defined as $d_{\Bty}(v_1, v_2) \leq (\senv[1] \sjoin \senv[2]) \cdot \denv$.

Since we have re-located the responsibility of introducing resources from lambdas to distance abstractions, the case for related functions is simpler than its \LamS counterpart (Figure~\ref{fig:basic-static-logical-relation}); related arguments produce related results, without the need to eliminate any variable from the types.
In turn, the case for related type abstractions considers the sensitivity environment argument, and eliminates the distance variable from the type.
Finally, related computations now consider the possibility of errors during evaluation: if the two expressions reduce to values, then they should be related according to the definition of related base values, otherwise the relation holds vacuously, similar to the definition of $\EMP$.
All other aspects are analogous to the ones presented in \S\ref{sec:sensitivity-typing-background}.

We define semantic gradual sensitivity typing just as before:

\begin{definition}[Semantic Gradual Sensitivity Typing]
\label{def:gradual-semantic-sensitivity-typing}
\begin{small}
\begin{align*}
  \semtypingGMP[\tenv]{e}{\cgty} \quad \defeq \quad {}
  & \forall \denv, \venv[1], \venv[2] \ST  (\venv[1], \venv[2]) \in \glrCoreG{\tenv}
  . {} \\
  & \quad  ({\venv[1](e)}, {\venv[2](e)}) \in \glrCoreT{\cgty}
\end{align*}
\end{small}
\end{definition}

Given that in the gradual setting we type expressions under plausibility-based constraints, soundness of semantic typing cannot be established for all possible sensitivities within the ranges of a gradual sensitivity environment $\gsenv$.
Nevertheless, the observed output distance should still be bounded by $\gsenvp \cdot \denv$, for some fully-precise $\gsenvp$ that is more precise than $\gsenv$.
We define $\GMP$, $\EMP$ for gradual sensitivities, as follows:

\begin{definition}[Gradual Metric Preservation]
\label{def:gradual-metric-preservation}
Let ${\tenv = x_1 : \itype{\Bty_1}{\gsenv[1]}, \dots, x_n : \itype{\Bty_n}{\gsenv[n]}}$.
An open expression $e$ satisfies gradual metric preservation, written $\GMP(\tenv, e, \Bty, \gsenv)$, iff for all $\denv$ and $\venv[1], \venv[2]$ such that $\venvClose[\denv]{\venv[1]}{\venv[2]}$:

\vspace*{-\baselineskip}

\begin{small}
\begin{align*}
  &(\reval{\venv[1](e)}{v_1} \land
    \reval{\venv[2](e)}{v_2}) \implies \\
  & \quad \exists \gsenvp \ST \mathsf{static}(\gsenvp) \land \gsenvp \gprec \gsenv .
  d_{\Bty}(v_1, v_2) \leq \gsenvp \cdot \denv
\end{align*}
\end{small}
\end{definition}

The predicate $\mathsf{static}(\gsenv)$ holds when all gradual sensitivities in $\gsenv$ are fully precise.
We then prove that the semantic typing is sound with respect to the gradual metric preservation:

\begin{lemma}[Semantic Gradual Sensitivity Typing is Sound]
\label{lm:gradual-semantic-sensitivity-soundness}
Let $\tenv = x_1 : \itype{\Bty_1}{\gsenv[1]}, \dots, x_n : \itype{\Bty_n}{\gsenv[n]}$.
\begin{small}
\begin{align*}
  \semtypingZ[\tenv]{e}{\itype{\Bty}{\gsenv}} \implies
    \GMP(\tenv, e, \Bty, \gsenv)
\end{align*}
\end{small}
\end{lemma}

\subsection{Termination-sensitive metric preservation}
\label{sec:tsmp}
\begin{figure*}
\begin{small}
\begin{align*}
  %
  (v_1, v_2) \in \tsglrCoreV{\itype{\Bty}{\gsenv}} \quad \defeq \quad {}
  &
  \getMon(v_1) \equiprec \getMon(v_2) \land
  v_i \in \llbracket \Bty \rrbracket \land
  d_{\Bty}(v_1, v_2) \leq (\senv[1] \sjoin \senv[2]) \cdot \denv
  \quad \text{\textcolor{black}{where}}~ \senv[i] = \getS(\getMon(v_i))\\
  %
  (v_1, v_2) \in \tsglrCoreV{\itype{(\cgty[1] \rightarrow \cgty[2])}{\gsenv}} \quad \defeq \quad {}
  &
  \getMon(v_1) \equiprec \getMon(v_2) \land
  \forall \svp[1], \svp[2], (v_1', v_2') \in \tsglrCoreV{\cgty[1]}.
  ({\sv[1] \app \svp[1]}, {\sv[2] \app \svp[2]}) \in \tsglrCoreT{\cgty[2] \effOp{+} \gsenv}\\
  (v_1, v_2) \in \tsglrCoreV{\itype{(\polyST.\cgty)}{\gsenv}} \quad \defeq \quad {}
  &
  \getMon(v_1) \equiprec \getMon(v_2) \land
  \forall \gsenvpp \ST \bounded(\gsenvpp). ({\sv[1] \appS{\gsenvpp}}, {\sv[2] \appS{\gsenvpp}}) \in \tsglrCoreT{\Xisubst{\cgty}[\r]{\gsenvpp} \effOp{+} \gsenv}\\
  %
  (e_1, e_2) \in \tsglrCoreT{\cgty} \quad \defeq \quad {} &
  \reval{e_1}{v_1} \implies \reval{e_2}{v_2} \land (v_1, v_2) \in \tsglrCoreV{\cgty}\\
  (\venv[1], \venv[2]) \in \tsglrCoreG{\tenv} \quad \defeq \quad {} &
  \bounded(\denv) \land \forall x \in \dom(\Gamma). (\venv[1](x),\venv[2](x)) \in \tsglrCoreV{\tenv(x)}
\end{align*}
\end{small}
  \caption{Logical relations for semantic termination-sensitive gradual sensitivity typing}
  \label{fig:gradual-logical-relations-ts}
\end{figure*}

The notion of gradual metric preservation presented above (Def.~\ref{def:gradual-metric-preservation}) is {\em termination insensitive}:
if either of the two executions fail, the property vacuously holds. This property is weaker than the one originally proposed by \citet{reedAl:icfp2010} for their static sensitivity type system. We now study how to achieve a stronger, {\em termination sensitive} characterization of gradual metric preservation, \ie~either both executions fail or both reduce to values at a bounded distance. We show that such a stronger property is achievable if the imprecision of sensitivities is bounded (\ie~not infinite).

A first observation towards enforcing the same termination behavior across runs is to realize that two substitutions $\venv[1]$ and $\venv[2]$ may introduce values with different levels of precision that may cause an error in only one of the executions.
To prevent this, we take inspiration on prior work~\cite{toroAl:popl2019} and restrict the monitors of related values to be \emph{equi-precise}, denoted by $\equiprec$.
Suppose that $m_i = \getMon(v_i)$, then $m_1 \equiprec m_2$ if and only if $m_1 \gprec m_2$ and $m_2 \gprec m_1$.\footnote{
  Notice that the ability to compare monitors under a precision relation is only necessary for the termination-sensitive logical relations.
}
Second, we draw from \citet{dantoniAl:fpdcsl2013} and \citet{azevedoDeAmorimAl:popl2017}, who show, in their respective calculi, that when restricting the predicted output distance to be finite, the termination behavior of executions for a given expression is uniform.
We define the following predicate to check if a distance environment (or a sensitivity environment) is bounded, \ie~does not contain $\SG{\infty}$:

\vspace*{-\baselineskip}

\begin{small}
\begin{align*}
  \bounded(\denv) \quad \defeq \quad \forall \r \in \dom(\denv). \SG{\infty} \not\gprec \denv(\r)
\end{align*}
\end{small}
Figure~\ref{fig:gradual-logical-relations-ts} presents the termination-sensitive logical relations for \GSoulCore types, denoted $\tsglrCoreV{\cgty}$, $\tsglrCoreT{\cgty}$, $\tsglrCoreG{\tenv}$.
The main differences with respect to their termination-insensitive counterparts are: (1) all definitions of related values now include a check for the equi-precision of their monitors, (2) related computations require the same termination behavior, and (3) related substitutions require the inputs distance to be finite.
Additionally, the case for related distance abstractions now requires the sensitivity environment argument to be bounded, \ie~$\bounded(\gsenvpp)$.
Then, we define semantic termination-sensitive typing as follows:

\begin{definition}[Semantic (Termination-Sensitive) Gradual Sensitivity Typing]
\label{def:ts-gradual-semantic-sensitivity-typing}
\begin{small}
\begin{align*}
  \semtypingGMPts[\tenv]{e}{\itype{\gty}{\gsenv}} \quad \defeq \quad {}
  & \bounded(\gsenv) \land {} \\
  & \forall \denv, \venv[1], \venv[2] \ST (\venv[1], \venv[2]) \in \tsglrCoreG{\tenv} 
  . {} \\
  & \quad  ({\venv[1](e)}, {\venv[2](e)}) \in \tsglrCoreT{\itype{\gty}{\gsenv}}
\end{align*}
\end{small}
\end{definition}

Besides relying on the termination-sensitive logical relations, Definition~\ref{def:ts-gradual-semantic-sensitivity-typing} also requires that the top-level sensitivity environment is bounded.
This latter requirement, in conjunction with the condition over $\denv$ in the definition of related substitutions, ensures that the output distance is not plausibly infinite, obtaining the same termination behavior across runs.
Moreover, notice that this criterion is modular, \ie~$e$ may use any form of imprecision internally, as long as the top-level sensitivity environment is bounded.
This definition illustrates how the progressive hardening of precision may allow a programmer to obtain stronger guarantees about their programs.

Finally, following the same ideas, we can define a termination-sensitive variant of $\GMP$:

\begin{definition}[(Termination-Sensitive) Gradual Metric Preservation]
\label{def:ts-gradual-metric-preservation}
Let $\tenv = x_1 : \itype{\Bty_1}{\gsenv[1]}, \dots, x_n : \itype{\Bty_n}{\gsenv[n]}$.
An open expression $e$ satisfies termination-sensitive gradual metric preservation, written $\GMP(\tenv, e, \Bty, \gsenv)$, iff for all $\denv$ and $\venv[1], \venv[2]$ such that $\Gbox{\venvCloseTS[\denv]{\venv[1]}{\venv[2]}}$:

\vspace*{-\baselineskip}

\begin{small}
\begin{align*}
  \reval{\venv[1](e)}{v_1} \implies {}& \reval{\venv[2](e)}{v_2} \land {} \\
  & \exists \gsenvp \ST \mathsf{static}(\gsenvp) \land \gsenvp \gprec \gsenv .
  d_{\Bty}(v_1, v_2) \leq \gsenvp \cdot \denv
\end{align*}
\end{small}
\end{definition}


The new definition of substitutions proximity, $\venvCloseTS{\cdot}{\cdot}$, analogous to $\glrCoreG{\tenv}$, requires $\denv$ to be bounded and related values to have equi-precise monitors.
Then, we have the following lemma:

\begin{lemma}[Soundness of (Termination-Sensitive) Gradual Semantic Typing]
\label{lm:ts-gradual-semantic-sensitivity-soundness}
Let $\tenv = x_1 : \itype{\Bty_1}{\gsenv[1]}, \dots, x_n : \itype{\Bty_n}{\gsenv[n]}$.

\vspace*{-\baselineskip}

\begin{small}
\begin{align*}
  \semtypingGMPts[\tenv]{e}{\itype{\Bty}{\gsenv}} \implies
    \GMPts(\tenv, e, \Bty, \gsenv)
\end{align*}
\end{small}
\end{lemma}

To conclude this section, we put our attention on the runtime semantics of \GSoulCore, which we have left abstract so far.
We know that the runtime of \GSoulCore must be able to monitor the sensitivity of expressions during their evaluation, and use this information to both perform runtime checks and provide a sound approximation of the actual sensitivity of the expressions.
In the next sections, we tackle this specification in two steps: in \S\ref{sec:gscore-runtime}, we present an intermediate language, variant of \GSoulCore, that is sound with respect to the semantic typing defined in this section; and, in \S\ref{sec:gscore-to-gscore-ev}, we define a syntax-directed and type-preserving translation from \GSoulCore to this intermediate language, which also plays the role of a syntactic type system for \GSoulCore.

\section{Runtime Semantics for \GSoulCore}
\label{sec:gscore-runtime}

To define the runtime semantics of \GSoulCore, we follow the general recipe of gradual languages, where source language expressions are statically typechecked and elaborated to an internal language with explicit runtime checks that account for the optimism that arises from imprecise type information~\cite{siekTaha:sfp2006,siekAl:snapl2015,garciaAl:popl2016}.
Specifically, we follow the general approach of \citet{garciaAl:popl2016}, which refines the initial justification of optimistic judgments in expressions as they reduce, yielding {\em type-driven} monitors for gradual programs. We first describe this approach in our setting of imprecise sensitivity information in \S\ref{sec:evidence}, and then define the internal language \GSoulEvCore in \S\ref{sec:reduction-rules}. \S\ref{sec:gscore-to-gscore-ev} then describes the syntactic typechecking and elaboration of \GSoulCore to \GSoulEvCore, and proves that any syntactically well-typed \GSoulCore term elaborates to a semantically well-typed \GSoulEvCore term that therefore satisfies (possibly termination-sensitive) metric preservation, as well as the properties expected from gradual languages~\cite{siekAl:snapl2015}.


\subsection{Evidence for Consistent Sensitivity Subtyping}
\label{sec:evidence}

To systematically gradualize typing disciplines, the Abstracting Gradual Typing approach~\cite{garciaAl:popl2016} recognizes that, in the presence of consistent judgments---subtyping, in our case---evaluation must be concerned about not only \emph{that} an optimistic judgment holds, but also \emph{why} it holds, locally. During reduction, these local justifications need to be combined to keep justifying the plausibility of subsequent steps, or fail otherwise to report a violation of the optimistic assumptions.

To illustrate, suppose we have $\semtypingGMP[\emptyenv]{e}{\itype{\rtype}{\SG{2}\r}}$ and we want to evaluate $e' = e \ascr{\itype{\rtype}{\unknown\r}} \ascr{\itype{\rtype}{\SG{1}\r}}$. Observe that (i)~each ascription in $e'$ is {\em locally justified} thanks to the use of $\unknown\r$ in the first ascription, and
(ii)~evaluating $e'$ should produce an error at runtime, given that the sensitivity of $e$ is $\SG{2}\r$, not $\SG{1}\r$.

In order to achieve this behavior, we must refute the transitivity of $\itype{\rtype}{\SG{2}\r} \ccsub \itype{\rtype}{\unknown\r}$ and $\itype{\rtype}{\unknown\r} \ccsub \itype{\rtype}{\SG{1}\r}$.
However, given that each judgment holds individually, this is only possible if we know that the actual sensitivity cannot be \emph{simultaneously} greater than 2 and less than 1: we care about the individual proofs of the consistent subtyping judgments and how they can be combined (or not) to justify their transitivity. To this end,
\citet{garciaAl:popl2016} introduce the notion of \emph{evidences} $\ev$ that represent the proofs of consistent subtyping judgments.
We write $\ev \evjust \cgty[1] \ccsub \cgty[2]$ to denote that the evidence $\ev$ {\em justifies} that $\cgty[1]$ is a consistent subtype of $\cgty[2]$.
During reduction, one can compute whether the transitivity of two consistent subtyping judgments is justified using {\em consistent transitivity} (noted $\transsub$) of their respective evidences:
given $\ev[1] \evjust \cgty[1] \ccsub \cgty[2]$ and $\ev[2] \evjust \cgty[2] \ccsub \cgty[3]$,
if $\ev[1] \transsub \ev[2]$ is defined, then $\ev[1] \transsub \ev[2] \evjust \cgty[1] \ccsub \cgty[3]$. If the combination is undefined, then a runtime error must be raised, as the transitivity has been effectively refuted.

We now formally define evidences and their operations.

\paragraph{Evidences}
In the presence of consistent subtyping~\cite{garciaAl:popl2016,toroAl:toplas2018}, an evidence $\ev$ is typically represented as a pair of \typeclsPlur, written $\evi{\cgty[1], \cgty[2]}$, each of which being at least as precise as the types involved in the consistent subtyping relation. Intuitively, the types in the evidence can be seen as refined versions of the types in the consistent subtyping judgment, each of them a proof that the judgment holds.


\paragraph{Interior}
Evidence is initially computed from a consistent subtyping judgment using the \emph{interior} operator, which essentially produces a refined pair of gradual types, based solely on the knowledge that the judgment holds.
Naturally, if the judgment does not hold, the interior operator is undefined.

The interior for sensitivity subtyping is naturally defined as the lifting of the interior for consistent sensitivity ordering:

\begin{definition}[Interior]
  \label{def:interior}
  \begin{align*}
    \Isub(\sensint{\sens[1], \sens[2]}, \sensint{\sens[3], \sens[4]}) =
    \evi{\sensint{\sens[1], \sens[2] \smeet \sens[4]}, \sensint{\sens[1] \sjoin \sens[3], \sens[4]}}
  \end{align*}
  If $\sens[1] \leq \sens[2] \smeet \sens[4]$ and $\sens[1] \sjoin \sens[3] \leq \sens[4]$.
  Otherwise, the interior is undefined.
\end{definition}
To illustrate, suppose we have $\cgty[1] = \teff{\rtype}{\unknown\r}$ and $\cgty[2] = \teff{\rtype}{10\r}$, noting that $\cgty[1] \ccsub \cgty[2]$.
We compute the initial evidence for that judgment as $\Isub(\cgty[1], \cgty[2]) = \evi{\teff{\rtype}{\sensint{0,10}\r}, \teff{\rtype}{\sensint{10,10}\r}}$.
Notice how the evidence is (component-wise) more precise than the types involved in the consistent subtyping judgment.
In particular, the range $\sensint{0,\infty}$ (corresponding to $\unknown$) gets refined to $\sensint{0,10}$, based on the knowledge that it is consistently inferior to $\sensint{10,10}$. The plausibility induced by $\unknown$ has been made more precise by excluding definitely invalid candidate sensitivities.

For a judgment $\ev \evjust \cgty[1] \ccsub \cgty[2]$ to be valid, the evidence must be at least as precise as the interior, \ie~$\ev \gprec^2 \Isub(\cgty[1], \cgty[2])$---where $\gprec^2$ is the precision relation checked component-wise, \ie~$\evi{\cgty[1], \cgty[2]} \gprec^2 \evi{\cgtyp[1], \cgtyp[2]} \iff \cgty[1] \gprec \cgtyp[1] \land \cgty[2] \gprec \cgtyp[2]$.

\paragraph{Consistent transitivity}

Finally, the {consistent transitivity operator} for consistent sensitivity ordering is defined as:

\begin{definition}[Consistent Transitivity]
  \label{def:ctrans}
  \begin{align*}
    &\evi{\sensint{\sens[11], \sens[12]}, \sensint{\sens[13], \sens[14]}} \transsub
    \evi{\sensint{\sens[21], \sens[22]}, \sensint{\sens[23], \sens[24]}} = \\
    &\evi{
      \sensint{\sens[11], \sens[12] \smeet \sens[14] \smeet \sens[22]},
      \sensint{\sens[13] \sjoin \sens[21] \sjoin \sens[23], \sens[24]}
    }
  \end{align*}

  If  $\sens[11] \leq (\sens[12] \smeet \sens[14] \smeet \sens[22])$ and
      $(\sens[13] \sjoin \sens[21] \sjoin \sens[23]) \leq \sens[24]$.
  Otherwise, the consistent transitivity operation is undefined.
\end{definition}

Same as with the interior, consistent transitivity is naturally defined for types, and is in the supplementary material.
%






\begin{figure*}[!t]
  \begin{minipage}{0.4\textwidth}
  \begin{small}
    \begin{align*}
      u ::={}&
        c \mid
        \lambda (x:\gty). e \mid
        \polyS. v
        \\
      v ::={}& \ev u \ascr{\cgty}
      %
      \end{align*}
  \end{small}
  \end{minipage}
  \begin{minipage}{0.58\textwidth}
  \begin{small}
    \begin{align*}
      e ::={}&
        v \mid
        \bop \overline{e} \mid
        x \mid
        e \app e \mid
        \ev (\polyS . e) \ascr{\cgty} \mid
        e \appS{\gsenv} \mid
        \ev e \ascr{\cgty}
        \\
      E ::={}&
        \emptycontext \mid \bop(\overline{v_i}, \Ectx, \overline{e_j}) \mid \Ectx \app e \mid v \app \Ectx \mid \Ectx \appS{\gsenv} \mid \ev (\polyS[\r] . \Ectx) \ascr{\cgty} \mid \ev \Ectx \ascr{\cgty}
    \end{align*}
  \end{small}
  \end{minipage}


    \begin{small}

    \begin{align*}
      %
      %
      \IGrefr{r-op} & &
      \bop \overline{(\ev[i] c_i \ascr{\cgty[i]})}
      &\longrightarrow
      \bopFuncName^2(\overline{\ev[i]}) (\mathbin{\llbracket \bop\! \rrbracket}\, \overline{c_i}) \ascr{\bopFuncName(\overline{\cgty[i]})}\\
      %
      %
      \IGrefr{r-app} & &
      (\ev (\lambda (x: \cgtyp[1]).e) \ascr{\cgty}) \app (\ev[1] u \ascr{\cgty[1]})
      &\longrightarrow
      \begin{cases}
        \icod(\ev)
          (\substv{(\ev[1] \transsub \idom(\ev)) u \ascr{\cgtyp[1]}}{x}{e})
         \ascr{\codd(\cgty)} & \\
        \error \hspace{1.7cm} \text{\color{black} if not defined}\\
      \end{cases} \\
      %
      %
      \IGrefr{r-inst} & &
      (\ev (\polyS[\r].v) \ascr{\cgty}) \appS{\gsenv}
      & \longrightarrow
      \iinst(\ev, \gsenv)
      (\substv{\gsenv}{\SG{\r}}{v})
      \ascr{\instt(\cgty, \gsenv)}\\
      %
      %
      \IGrefr{r-ascr} & &
      \ev (\evp u \ascr{\cgtyp}) \ascr{\cgty}
      & \longrightarrow
      \begin{cases}
        (\evp \transsub \ev) u \ascr{\cgty} & \\
        \error & \text{\color{black}if not defined}
      \end{cases}
    \end{align*}

    \begin{align*}
      &\domm(\itype{(\cgty[1] \rightarrow \cgty[2])}{\gsenv}) = \cgty[1] &
      &\codd(\itype{(\cgty[1] \rightarrow \cgty[2])}{\gsenv}) = \cgty[2] \effOp{+} \gsenv
      &
      &\instt(\itype{(\polyS.\cgty[2])}{\gsenv}, \gsenv[1]) = \substv{\gsenv[1]}{\SG{\r}}{\cgty[2]} \effOp{+} \gsenv \\
      &\idom(\evi{\cgty[1], \cgty[2]}) = \evi{\domm(\cgty[2]), \domm(\cgty[1])} &
      &\icod(\evi{\cgty[1], \cgty[2]}) = \evi{\codd(\cgty[1]), \codd(\cgty[2])} &
      &\iinst(\evi{\cgty[1], \cgty[2]}, \gsenv) = \evi{\instt(\cgty[1], \gsenv), \instt(\cgty[2], \gsenv)}
    \end{align*}

  \end{small}



  \caption{Dynamic semantics of \GSoulEvCore}
  \label{fig:gscoreev-reduction}
\end{figure*}

\subsection{Evidence-Based Reduction}
\label{sec:reduction-rules}
We now present \GSoulEvCore, the internal language for \GSoulCore, which it enriches with evidences.
Figure~\ref{fig:gscoreev-reduction} shows the syntax and notions of reduction (denoted by the use of $-->$) for \GSoulEvCore.
Syntactically, an ascription is augmented with evidence that justifies its intrinsic consistent subtyping judgment. Values $v$ correspond to ascribed simple values $u$, and introduction forms are also ascribed (this syntactic uniformity minimizes the cases in reduction rules).

\paragraph{Reduction}
Rule \IGrefr{r-op} relies on the metafunction $\llbracket \bop\! \rrbracket$ and the resulting type is computed using the operator $\bopFuncName$.
This operator is generalized to work with evidences, and the resulting evidence is computed using the $\bopFuncName^2$ metafunction, which is defined as
$\bopFuncName^2(\overline{\evi{\cgty[i1], \cgty[i2]}}) =
\evi{
  \bopFuncName(\overline{\cgty[i1]}),
  \bopFuncName(\overline{\cgty[i2]})
}$.

The application rule \IGrefr{r-app} (1) ascribes the argument to the expected type $\cgtyp[1]$, computing a proper evidence by combining the original evidence with the domain part of the evidence of the lambda; and, (2) ascribes the body of the lambda to the expected codomain type, using the codomain part of the evidence of the lambda.
Inversion metafunctions $\idom$ and $\icod$ are used to extract the domain and codomain parts of the evidence, respectively.
If the combination of the evidences is not defined, the rule halts with an error.
This treatment guarantees that all expected sensitivity types are respected during runtime.

Rule \IGrefr{r-inst} reduces the application of a reduced resource abstraction simply by substituting the distance variable $\r$ by the provided sensitivity environment $\gsenv$.
The body of the abstraction, analogous to the application rule, is ascribed to the instantiated type, which is computed using the $\instt$ meta-function.
The body of a distance abstraction $\polyS[\r] . e$ is reduced before the substitution is performed.
\footnote{
If we simply substitute $\r$ in $e$, we may lose information necessary to refute optimistic static assumptions, failing to produce an error at runtime.
A minimal example is the expression $(\polyS[\r[2]]. c \ascr{\r[1]} \ascr{\unknown\r[1]} \ascr{\r[2]}) \appS{\r[1]}$, where $c$ is a constant and $\r[1]$ is in scope.
This is reminiscent of a similar issue in gradual parametricity~\cite{ahmedAl:icfp2017,toroAl:popl2019}.}%
%

Finally, rule \IGrefr{r-ascr} merges ascriptions by keeping only the outer type $\cgty$ and using consistent transitivity to attempt to justify that the type of $u$ is a consistent subtype of $\cgty$. If consistent transitivity is undefined, the program halts with an error.

The reduction of \GSoulEvCore, denoted $\cred{e}{e'}$, is call-by-value, defined with evaluation contexts $\Ectx$ and given by the following rules:

\vspace*{-\baselineskip}

\begin{small}
\begin{mathpar}
\inferrule*[left=\IGrefr{R$\Ectx$}]{
    e_1
    \longrightarrow
    e_2
}{
  \cred[]{
    \Ectx[][e_1]
  }{
    \Ectx[][e_2]
  }
} \and
\inferrule*[left=\IGrefr{R$\Ectx$err}]{
  e_1 \longrightarrow \error
}{
  \cred[]{E[e_1]}{\error}
}
\end{mathpar}
\end{small}
An evaluation context $\Ectx$ may be seen as a meta-expression with a hole, where the hole can be filled by an expression $e$, written as $\Ectx[][e]$.
The use of evaluation contexts allows us to decouple the congruence rules, \eg~the order of reduction of operands, from the actual notions of reduction, making the presentation more concise.
Finally, reduction $\reval{e}{v}$---as used in Figure~\ref{fig:gradual-logical-relations}---is defined as $\creds{e}{v}[*]$.

\paragraph{Evidences as Runtime Monitors}
In order to connect \GSoulEvCore and the general treatment of monitors used in the logical relations defined in \S\ref{sec:semantics-gscore-ev}, we need to instantiate the abstract operators $\getMon$ and $\getS$---thereby precisely establishing that evidences are the type-driven monitoring mechanism of \GSoulEvCore.
If an expression successfully reduces to an ascribed value $\ev u \ascr{\cgty}$, then the evidence $\ev$ is guaranteed to convey a sound conservative approximation of the sensitivity of the computation that produced $u$, in particular more precise than that contained in $\cgty$. Indeed, observe that consistent transitivity is monotone with respect to precision~\cite{garciaAl:popl2016}, tightening precision as reduction proceeds.
Formally, if $\creds{e}{\evi{\_,\cgtyp} u \ascr{\cgty}}[*]$, then $\cgtyp \gprec \cgty$.
To illustrate, suppose that we have $\creds{e}{\evi{\_, \itype{\rtype}{\sensint{a,b}\r}} c \ascr{\itype{\rtype}{\unknown\r}}}[*]$ (we omit the left-hand side of the evidence because we want to focus on the evidence counterpart of the ascription type).
By construction, evidence operations and reduction rules ensure that the interval $\sensint{a,b}$ is a conservative approximation of the runtime sensitivity of $e$, \ie~any sensitivity within the interval is a valid sensitivity bound for $e$.
In particular, the best \emph{monitored sensitivity} of the expression is obtained by taking the lower bound of the right-hand side of the evidence, \ie~$\SG{a}$.
Given this, we instantiate the abstract operators $\getMon$ and $\getS$ as follows:

\vspace*{-\baselineskip}

\begin{small}
\begin{align*}
  &\getMon(\ev u \ascr{\cgty}) = \ev \qquad  \getS(\evi{\_, \itype{\gty}{\gsenv}}) = \mathsf{lower}(\gsenv)\\
  &\text{\black where }\mathsf{lower}(\sensint{\sens[1], \_}\r[1] + \dots + \sensint{\sens[n], \_}\r[n]) = \sens[1]\r[1] + \dots + \sens[n]\r[n]
\end{align*}
\end{small}
\noindent The function $\mathsf{lower}$ returns a static sensitivity environment obtained from the lower bounds of each gradual sensitivity in its argument.

\section{Syntactic Typing and Elaboration}
\label{sec:gscore-to-gscore-ev}

\begin{figure*}[t]
\begin{small}
\begin{mathpar}
  \inferrule*[lab=\ELrefr{const}]{
    c \in \llbracket \Bty \rrbracket\\
    \cgty = \itype{\Bty}{\emptysenv}\\
    \ev = \Isub(\cgty, \cgty)
  }{
    \itelab{c}{\ev c \ascr{\cgty}}{\cgty}
  }

  \inferrule*[lab=\ELrefr{op}]{
    \overline{
      \itelab{\e[i]}{\ep[i]}{\cgty[i]}
    }\\
    \bopFuncName( \overline{\cgty[i]} ) = \cgty
  }{
    \itelab{
      \bop \overline{\e[i]}
    }{
      \bop \overline{\ep[i]}
    }{
      \cgty
    }
  }

  \inferrule*[lab=\ELrefr{var}]{
    \tenv(x) = \cgty
  }{
    \itelab{x}{x}{\cgty}
  }

  \inferrule*[lab=\ELrefr{lam}]{
    \itelab[\tenv, x:\cgty[1]]{\e}{\ep}{\cgty[2]}\\
    \cgty = \itype{(\cgty[1] \rightarrow \cgty[2])}{\emptysenv}\\
    \ev = \Isub(\cgty, \cgty)
  }{
    \itelab{
      \lambda (x: \cgty[1]).\e
    }{
      \ev (\lambda (x: \cgty[1]).\ep) \ascr{\cgty}
    }{
      \cgty
    }
  }

  \inferrule*[lab=\ELrefr{app}]{
    \itelab{\e[i]}{
      \ep[i]
    }{\cgty[i]}\\
    \ev[2] = \Isub(\cgty[2], \domm(\cgty[1]))
  }{
    \itelab{
      \e[1] \app \e[2]
    }{
      \ep[1] \app
      (\ev[2] \ep[2] \ascr{\cgty[1]})
    }{\codd(\cgty[1])}
  }

  \inferrule*[lab=\ELrefr{$\Lambda$},width=8cm]{
    \itelab[\tenv]{\e}{\ep}{\cgty}\\
    \cgty = \itype{(\polyST . \cgty)}{\emptysenv}\\
    \ev = \Isub(\cgty, \cgty)
  }{
    \itelab{
      \polyS . \e
    }{
      \ev \polyS . \ep \ascr{\cgty}
    }{
      \cgty
    }
  }

  \inferrule*[lab=\ELrefr{inst}]{
    \itelab{\e}{
      \ep
    }{\cgty}
  }{
    \itelab{
      \e \appS{\gsenv}
    }{
      \ep \appS{\gsenv}
    }{\instt(\cgty, \gsenv)}
  }

  \inferrule*[lab=\ELrefr{ascr}]{
    \itelab{\e}{\ep}{\cgty}\\
    \ev = \Isub(\cgty,\cgtyp)
  }{
    \itelab{\e \ascr{\cgtyp}}{
      \ev \ep \ascr{\cgtyp}
    }{\cgtyp}
  }



\end{mathpar}

\end{small}

\caption{Syntactic Elaboration for \GSoulCore}
\label{fig:gscore-elaboration}
\end{figure*}

Armed with the definition of \GSoulEvCore, we can now establish the main results of this work: a syntactic type system for \GSoulCore that ensures that well-typed expressions satisfy gradual metric preservation. Technically, we define a syntactic type-directed elaboration from \GSoulCore to \GSoulEvCore (\S\ref{sec:elaboration}). This elaboration plays two roles: it is an algorithmic syntactic typechecking procedure for \GSoulCore, and it produces terms
from the internal language \GSoulEvCore that can then be executed.
We prove that if a \GSoulCore expression is successfully elaborated---\ie~that it is syntactically well-typed---then the produced \GSoulEvCore expression is semantically well-typed.
Finally, we prove that \GSoulCore satisfies standard type safety as well as the gradual guarantees~\cite{siekAl:snapl2015}, and crucially, gradual sensitivity type soundness (\S\ref{sec:metatheory}).

\subsection{Type-Directed Elaboration}
\label{sec:elaboration}

Figure~\ref{fig:gscore-elaboration} shows the type-directed elaboration from \GSoulCore to \GSoulEvCore.
Rules \ELrefr{const}, \ELrefr{lam} and \ELrefr{$\Lambda$} generate evidences for ascribing introduction forms using the interior operator, $\Isub$.
Notice that the interior is always defined in these rules, since the types are equal, hence in the consistent subtyping relation.
For \ELrefr{lam} and \ELrefr{$\Lambda$}, the bodies of lambdas and distance abstractions are elaborated inductively.

Rule \ELrefr{op} elaborates the arguments of an operator application and computes the type using the operator $\bopFuncName$ (similar to the typing rule presented in Figure~\ref{fig:lambdaS-syntactic-typing}).
Rule \ELrefr{var} simply retrieves the type of the variable from the typing environment.

Rule \ELrefr{app} inductively elaborates both inner expressions, but also ascribes the argument of the application to the exact type expected by the callee.
Consequently, an evidence for such ascription is produced using the interior operator.
Note however that the interior may not be defined for the involved types, thereby revealing a syntactic type error.

Rule \ELrefr{inst} simply elaborates the callee and computes the type using the $\instt$ function.
Finally, rule \ELrefr{ascr} is straightforward, as it simply elaborates the inner expression and ascribes it to the specified type. If the interior is not defined, a syntactic type error is exhibited.

Crucially, if elaboration succeeds, \ie~the source expression is syntactically well-typed, then the elaborated expression is semantically well-typed---possibly termination-sensitive if its gradual sensitivity is bounded.

\begin{lemma}[Syntactic Elaboration implies Semantic Typing]
\label{lm:elaboration-implies-semtyping}
If $\itelab[\tenv]{e}{e'}{\itype{\gty}{\gsenv}}$ then:

\begin{enumerate}
  \item $\semtypingGMP[\tenv]{e'}{\itype{\gty}{\gsenv}}$, and
  \item if $\bounded(\gsenv)$ then  $\semtypingGMPts[\tenv]{e'}{\itype{\gty}{\gsenv}}$
\end{enumerate}
\end{lemma}

Finally, by means of the elaboration, we can define a syntactic typechecking judgment and reduction for \GSoulCore expressions:


\vspace*{-\baselineskip}

\begin{small}
\begin{align*}
  \syntypingGMP{e}{\cgty} &\ \defeq \ \itelab{e}{e'}{\cgty} \text{ \black for some } e' \\
  \reval{e}{r} &\ \defeq \ {} \itelab[\emptytenv]{e}{e'}{\cgty} \land \creds{e'}{r}[*]
\end{align*}
\end{small}
In the evaluation judgment $\reval{e}{r}$, $e$ is a \GSoulCore expression and the result $r$ is either a \GSoulEvCore value $v$ or $\error$.

%% file: sections/metatheory.tex
\subsection{Metatheory of \GSoulCore}
\label{sec:metatheory}

We now establish that \GSoulCore is type safe, satisfies the gradual guarantees, and is type sound in that syntactically well-typed terms satisfy gradual metric preservation.
Note that some of these results follow from the results for the internal evidence-based language combined with the fact that syntactic typing for \GSoulCore is directly defined as the successful elaboration to \GSoulEvCore.

First, \GSoulCore is type safe: closed expressions do not get stuck. 

\begin{restatable}[Type Safety]{proposition}{gradualTypeSafety}
  \label{prop:gradual-type-safety}
  \reviv{Let $e$ be a closed \GSoulCore expression.}
  If
  $\syntypingGMP[\emptytenv]{e}{\cgty}$, then $\reval{e}{r}$. 
\end{restatable}

\GSoulCore satisfies the static and dynamic gradual guarantees~\cite{siekAl:snapl2015}.
The static guarantee says that typeability is monotone with respect to imprecision.\footnote{Precision between expressions is the standard, natural syntactical lifting of type precision to terms (provided in the supplementary material).}

\begin{restatable}[Static Gradual Guarantee]{proposition}{sgg}
  \label{prop:lifted-sgg}
  Let $e_1$ and $e_2$ be two closed \reviv{\GSoulCore} expressions such that $e_1 \gprec e_2$ and $\syntypingGMP[\emptytenv]{e_1}{\cgty[1]}$.
  Then, $\syntypingGMP[\emptytenv]{e_2}{\cgty[2]}$ and $\cgty[1] \gprec \cgty[2]$.
\end{restatable}

\GSoulCore also satisfies the dynamic gradual guarantee: any program that reduces without error will continue to do so if imprecision is increased.

\begin{restatable}[Dynamic Gradual Guarantee]{proposition}{dgg}
  \label{prop:dgg}
  \reviv{Let $e_1$ and $e_2$ be two well-typed closed \GSoulCore expressions such that } $e_{1} \gprec e_{2}$.
  If $\reval{e_1}{v_1}$
  then $\reval{e_2}{v_2}$,
  where $v_1 \gprec v_2$.
\end{restatable}


Any syntactically well-typed \GSoulCore term satisfies gradual metric preservation. More specifically, a well-typed term satisfies termination-insensitive gradual metric preservation, and if its type has bounded imprecision, then it also satisfies termination-sensitive gradual metric preservation.

\begin{theorem}[Sensitivity Type Soundness]
\label{th:gradual-soundness}
Let ${\tenv = x_1 : \itype{\Bty_1}{\gsenv[1]}, \dots, x_n : \itype{\Bty_n}{\gsenv[n]}}$.
If 
$\itelab[\tenv]{e}{e'}{\itype{\Bty}{\gsenv}}$ then:
\begin{enumerate}
  \item $\GMP(\tenv, e', \Bty, \gsenv)$, and
  \item if $\bounded(\gsenv)$ then $\GMPts(\tenv, e', \Bty, \gsenv)$
\end{enumerate}
\end{theorem}

%% file: sections/dp_reasoning.tex
\section{Differential Privacy Reasoning}
\label{sec:dp_reasoning}

Differential privacy~\cite{dworkRoth:fttcs2014} is a strong formal property that ensures that the output of a computation (or query) does not reveal information about any single element in the input data. Formally, a program $M$ is $\epsilon$-differentially private if for all neighboring inputs $x$ and $x'$, and for all possible outputs $r$ we have $\mathsf{Pr}[M(x) = r] \leq e^\epsilon \mathsf{Pr}[M(x') = r]$.
In general, differential privacy algorithms assume pure (total) functions, and hence
the possibility of leaks happening through abnormal termination behavior like errors is usually not considered.

Because sensitivity analysis is often used to build differentially-private programs, it is important to address the question of differential privacy built on top of a gradual sensitivity language like \lang.
In this section, we first illustrate why  {\em termination-insensitive} gradual metric preservation falls short of achieving privacy. We then demonstrate that {\em termination-sensitive} gradual metric preservation can be used in differential privacy proofs to ensure not only that sensitivity specifications are respected, but also that a program does not leak information through termination behavior. Specifically, we present some gradual variants of core differentially-private mechanisms and algorithms in \lang that leverage termination-sensitive metric preservation, which can serve as fundamental components in larger differentially-private programs.

\subsection{Termination Insensitivity and Differential Privacy}
We start by considering expressions that only satisfy termination-insensitive metric preservation (Definition~\ref{def:gradual-metric-preservation}).
In such a scenario, the differential privacy definition can be violated: it is possible to find a program $M$, such that given two different inputs $x$ and $x'$ at distance 1, then $\forall r. \mathsf{Pr}[M(x) = r] \not\leq e^\epsilon \mathsf{Pr}[M(x') = r]$. This occurs because $M(x')$ could result in an error (or diverge), reducing the probability $\mathsf{Pr}[M(x') = r]$ to $0$. To illustrate this, consider the following typing judgment:

\vspace*{-\baselineskip}

\begin{small}
\begin{align*}
  & x: \teff{\rtype}{\r}, f: \itype{(\itype{\rtype}{\r} \to \itype{\btype}{\unknown\r})}{\emptysenv} \\
  &\vdash \mathsf{if}\, (f \app x) \, \mathsf{then}\, 1 \,\mathsf{else}\, (x \ascr{\teff{\rtype}{?\r}} \ascr{\teff{\rtype}{0\r}}) : \teff{\rtype}{\unknown\r}
\end{align*}
\end{small}
The resulting type for this expression is computed by joining the types of the two branches of the conditional, $\teff{\rtype}{0\r}$ and $\teff{\rtype}{0\r}$, with the sensitivity environment of the condition, $\unknown\r$.
Note that this well-typed term fails at runtime for any $x$ and $f$ such that $f \app x$ evaluates to $\mathsf{false}$.
Then, in order for two executions to exhibit different termination behavior, it suffices to pick environments $\denv = \SG{1}\r$, and:

\vspace*{-\baselineskip}

\begin{small}
\begin{align*}
  \venv[1] &= \set{f \mapsto \lambda x. x > 0, x \mapsto 0} &
  \venv[2] &= \set{f \mapsto \lambda x. x > 0, x \mapsto 1}
\end{align*}
\end{small}
Importantly, this expression is not semantically well-typed in a termination-sensitive regime, as $\SG{\unknown}\r$ is not bounded.
If the codomain of $f$ were bounded, $x > 0$ could not be admitted as the body of $f$, and $\venv[1]$ and $\venv[2]$ could not be used to distinguish executions.
This observation highlights the importance of strengthening gradual programs so that imprecision is bounded in order to achieve differential privacy.


\subsection{Gradual Laplace Mechanism}

\begin{figure}

\begin{small}
\begin{lstlisting}[language=gsoul,numbers=left,xleftmargin=2em,escapechar={|}]
def GLM<T>(
      res x: T,
      f: T[|\color{StoreGreen}1x|] -> Number[|\color{StoreGreen}?x|],
      eps: Number,
) = laplace(f(x) :: Number[|\color{StoreGreen}1x|], 1, eps);
\end{lstlisting}
\end{small}

\caption{Gradual Laplace Mechanism (GLM)}
\label{fig:GLM}
\end{figure}

The Laplace mechanism is a fundamental building block in differential privacy, which adds noise drawn from the Laplace distribution to the output of a query.
In particular, if a query $Q$ is $\SG{1}$-sensitive, then $Q(D) + \mathcal{L}(1/\epsilon)$ is $\epsilon$-differentially private, for any database $D$~\cite{dworkRoth:fttcs2014}.
With gradual sensitivities, we define the \emph{Gradual Laplace Mechanism} (GLM) which expects an unknown-sensitive function and performs a runtime check to ensure that the function is $\SG{1}$-sensitive.
Figure~\ref{fig:GLM} presents the definition of GLM in \lang, parametric on the type (\lstinline|T|) of the input \lstinline|x|.


Given a function \lstinline|f|, the expression \lstinline[mathescape]|f(x) :: Number[$\SG{\text{1x}}$]| has a bounded sensitivity so it is termination-sensitive semantically well-typed.
Therefore, for any two values for \lstinline|x| (which are necessarily at a bounded distance), GLM either (1)~always fails for any input \lstinline|x|, or (2)~always produces noisy values and is \lstinline|eps|-differentially private.

\subsection{Gradual Above Threshold}

Recall the {Above Threshold} (AT) algorithm discussed in \secref{sec:in-action}.
This algorithm uses the Laplace mechanism as a subroutine to add noise to query results, which are then compared with a (noisy) threshold~\cite{dworkRoth:fttcs2014}.
We follow the same approach to define a gradual version of the AT algorithm, which instead uses the gradual Laplace mechanism GLM defined previously.
The new \emph{Gradual Above Threshold} (GAT) algorithm expects a list of queries of unknown sensitivities, but instead of failing if the sensitivity of a given function is greater than $\SG{1}$,
it simply skips the function.
Figure~\ref{fig:GAT} presents the GAT algorithm, written in \lang.
First, the algorithm computes a noisy version of the threshold (line 7).
Then, we use the \lstinline|indexOf| function to find the index of the first query whose (noisy) result is above the noisy threshold.
Notice that, given the use of GLM, executing line 11 may fail, but the \lstinline|try/catch| surrounding it ensures that in such case the algorithm continues with the next query.
In contrast to AT, GAT can receive a list of gradually well-typed functions of varied sensitivities, and the algorithm will only consider the $\SG{1}$-sensitive ones.
Hence, the algorithm returns the index of the \emph{first} $\SG{1}$-sensitive query whose result is above \lstinline|thr|.



\begin{figure}
\begin{small}
\begin{lstlisting}[language=gsoul,numbers=left,xleftmargin=2em,escapechar={|}]
def GAT<T>(
      res db: T,
      fs: List<T[|\color{StoreGreen}1db|] -> Number[|\color{StoreGreen}?db|]>,
      thr: Number,
      eps: Number,
): Number = {
  let noisyThr = laplace(thr, 1, eps / 2);
  fs.indexOf(
    fn (f: T[|\color{StoreGreen}1db|] -> Number[|\color{StoreGreen}?db|]) =>
      try {
        let noisyVal = GLM<T>(db, f, eps / 4);
        noisyVal >= noisyThr;
      } catch { false; }
  );
};
\end{lstlisting}
\end{small}

\caption{Gradual Above Threshold (GAT)}
\label{fig:GAT}
\end{figure}



%
The proof that GAT is differentially private follows the same methodology used by~\citet{dworkRoth:fttcs2014}.
The key difference is that, in the case of GAT, the proof first establishes a termination-sensitive result for the GLM call.
Thus, for a particular index, line 11 will always behave the same, either succeeding or failing.
This guarantees that the same functions are skipped every time, independently of the database.
Finally, given that the discarded functions do not reveal information about the database, the same proof structure of AT can be followed (see supplementary material).



Note that with termination-sensitive metric preservation, runtime consistent transitivity errors only reveal information about the sensitivity of functions, {\em not about their arguments}, which is sufficient to preserve differential privacy.
For instance, for GLM, an error will only reveal that the function passed as argument is not $\SG{1}$-sensitive, independent of the input value.
Proving that this holds in general, along with integrating this kind of reasoning in a differential privacy programming language such as \Fuzz~\cite{reedAl:icfp2010} or \Jazz~\cite{toroAl:toplas2023} is left to future work.

%% file: sections/related_work.tex
\section{Related Work}
\label{sec:related}

\paragraph{Sensitivity and programming languages}
The first type system for reasoning about sensitivity is \textsc{Fuzz}~\cite{reedAl:icfp2010}, a language for differential privacy using linear types.
Several variations have been studied, such as \textsc{DFuzz} \cite{gaboardiAl:popl2013}, \textsc{Fuzzi} \cite{zhangAl:icfp2019}, and \textit{Adaptive Fuzz} \cite{winogradcortAl:icfp2017}. $\color{black} \mu$\textsc{Fuzz} \cite{dantoniAl:fpdcsl2013} extends the \textsc{Fuzz} compiler to generate nonlinear constraints that are discharged by an SMT solver, resulting in an automatic type-based sensitivity analysis.
All of these type systems measure sensitivity and also track and enforce differential privacy.
\citet{nearAl:oopsla2010} tackle differential privacy in \textsc{Duet} with two mutually-defined languages, one dedicated to sensitivity and one to privacy.
\textsc{Jazz} \cite{toroAl:toplas2023} follows the approach of \textsc{Duet}, and includes \textsc{Sax}, a sensitivity language with contextual linear types and delayed sensitivity effects. The starting point of \lang, the static language \lambdaSens{}, is very close to \textsc{Sax}.
Whereas \textsc{Fuzz}-like languages track the sensitivity of program variables using linear types, \citet{abuahAl:arxiv2021a} propose \textsc{Solo}, which tracks a fixed amount of sensitive resources and avoids linear types.
\citet{loboAl:toplas2021} develop a haskell library, \textsc{DPella}, able to reason about the accuracy of differentially private queries.
\textsc{DDuo}~\cite{abuahAl2021b} is a library for dynamic sensitivity tracking in Python, which tackles the expressiveness issues of static sensitivity type systems.
However, as a fully dynamic approach, \textsc{DDuo} does not provide any static guarantee; in contrast, \lang allows programmers to balance static and dynamic checking as needed.
To the best of our knowledge, this integration of static and dynamic sensitivity analysis within one language is unique to \lang.

\begin{ppsalb}
Regarding termination sensitivity,
\Fuzz~\cite{reedAl:icfp2010} and more recent languages~\cite{gaboardiAl:popl2013,abuahAl:arxiv2021a}
focus on the terminating fragment of their languages.
\citet{reedAl:icfp2010} discuss the tension between metric preservation and non-termination in their paper, presenting three alternatives:
    weakening the definition of metric preservation;
    proving statically that recursive functions terminate, yielding more complex programs;
    or adding \emph{fuel} to recursive functions, falling back to a default value when exhausted. The latter is adopted in the implementation.
  \muFuzz \cite{dantoniAl:fpdcsl2013} also supports recursive types, and termination-sensitive metric preservation is obtained by reasoning only about finite distances.
  \textsc{Duet} \cite{nearAl:oopsla2010} avoids divergence via terminating looping primitives.
  \DDuo~\cite{abuahAl2021b} establishes termination-insensitive metric preservation, although they do not explicitly discuss this aspect.
\end{ppsalb}

Another line of work for sensitivity verification is based on program logics~\cite{bartheAl:popl2012,bartheAl:tpls2013,bartheAl:lics2016,satoAl:lics2019}.
These approaches are generally very expressive but less automatic than type systems.
Extending gradual verification approaches~\cite{baderAl:vmcai2018,wiseAl:oopsla2020}, to account for the aforementioned logics for sensitivity would be an interesting venue for future work.

\paragraph{Gradual typing}
To the best of our knowledge, gradual typing has not been applied to sensitivity typing.
It has, however, been applied in languages with one of two particularly interesting properties: languages with type-and-effect disciplines; and languages whose soundness property corresponds to a hyperproperty, such as noninterference~\cite{goguenMeseguer:ieee1982} or parametricity~\cite{reynolds:83}.
\ppsal{Interestingly, gradual security typing has only been explored for termination- (and error-)insensitive characterizations of noninterference~\cite{toroAl:toplas2018,fennellThiemann:csf2013,deAmorim:lics2020}.}
\hide{As mentioned previously, }\citet{banadosAl:icfp2014,banadosAl:jfp2016} develop a general approach to gradualize type-and-effects, extended by \citet{toroTanter:oopsla2015} to handle effect polymorphism in Scala.
Being based on the generic type-and-effect system of~\citet{marinoMillstein:tldi2009}, these approaches cannot handle sensitivities as quantities within a range.

\citet{toroAl:toplas2018} uncovered that the presence of mutable references in a security language yields a gradual language that does not satisfy noninterference; ad-hoc changes to address implicit flows recover noninterference, at the expense of the dynamic gradual guarantee.
An interesting perspective is to study an extension of \lang{} with mutable references, investigating if metric preservation and the dynamic gradual guarantee are both satisfied. They also use intervals (of security labels) in their runtime semantics in order to achieve soundness; \lang chooses to expose intervals in the source language for convenience. \citet{banados:arxiv2020} study \emph{forward completeness} of evidence representation as a criterion to ensure that modular type-based semantic invariants are preserved in a gradual language.
Proving forward completeness for \GSoulEvCore is an interesting exercise for future work.

%% file: sections/conclusion.tex
\section{Conclusion}
\label{sec:conclusion}

We have proposed gradual sensitivity typing as a means to seamlessly combine static and dynamic checking of sensitivity specifications, accommodating features and programming patterns that are challenging for fully static approaches. We designed and implemented \lang, a functional programming language that can be used to experiment with this approach. We formalized a core of \lang and studied its metatheory.

In particular, we have developed a semantic account of gradual metric preservation, taking into consideration its termination aspect, inevitable in a gradual language with runtime errors, and 
illustrated how to reason about differential privacy with gradual sensitivities.


This work represents an important first step towards gradual differentially-private programming. Future work will focus on building a privacy layer on top of \lang, following the two-language approach of \textsc{Duet}~\cite{nearAl:oopsla2010} and \textsc{Jazz}~\cite{toroAl:toplas2023}.
It is also crucial to further empirically evaluate the benefits and tradeoffs of gradual sensitivity typing, which requires further growing \lang to integrate practical libraries, or by implementing gradual sensitivities in a widely-used programming language.\\
